\documentclass[onecolumn,tightenlines,nofootinbib,preprintnumbers,amsmath,amssymb]{revtex4}
\usepackage[usenames,dvipsnames]{color}
\usepackage{caption}
\usepackage{dcolumn}
\usepackage{graphicx}  
\usepackage{caption} 
\usepackage{graphicx}
\usepackage{subfigure}
\usepackage{epstopdf}
\usepackage{bm}
\usepackage{lscape}
\usepackage{slashed}
\usepackage{booktabs}
\usepackage{appendix}
\renewcommand{\arraystretch}{0.9}
\usepackage{setspace}
\usepackage{makecell}

\begin{document}
\begin{spacing}{1.5}
\title{The role and contribution of resonance effect for the decay process of $\bar B^{0}_s \rightarrow \pi^{+}\pi^{-}P$}

\author{Xi-Liang Yuan$^{1}$\footnote{Email: xiliangyuan18@sina.com}, Chao Wang $^{1}$\footnote{Email: chaowang@nwpu.edu.cn}, Zhuang-Dong Bai $^{1}$\footnote{Email: baizhuangdong@163.com}, Gang L\"{u}$^{2 }$\footnote{Email: ganglv66@sina.com}}

\affiliation{\small $^{1}$School of Ecology and Environment, Northwestern Polytechnical University, Xi'an 710072, China\\
	\small $^{2}$ Institute of Theoretical Physics, College of Physics, Henan University of Technology, Zhengzhou 450001, China}

\begin{abstract}
The size of the direct CP asymmetry generated during the weak decay of hadrons is attributed to the weak phase and some strong phases. The weak phase comes from the CKM matrix and a strong phase may
result from the resonance effect which is produced by the mixing of vector meson $V\left\{\rho^{0}(770),\omega(782),\phi(1020)\right\}$ to $\pi^+ \pi^-$ meson pairs.
$\rho^{0}(770)$ can decay directly into $\pi^+ \pi^-$ meson pairs,
both $\omega(782)$ and $\phi(1020)$ can also decay into $\pi^+ \pi^-$ meson pairs with small contribution from isospin symmetry breaking. The main contribution for the middle state vector meson $\rho^{0}(770)-\omega(782)-\phi(1020)$ interference is the mix of $\rho^{0}(770)$, $\omega(782)-\rho^{0}(770)$ and $\phi(1020)-\rho^{0}(770)$.
We calculate the CP asymmetry and decay branching ratio for $\bar{B}^0_{s} \rightarrow \pi^+ \pi^- \pi^0 (\bar K ^{0})$ in the framework of QCD factorization and compare them with previous work. We also add the analysis of $\bar{B}^0_{s} \rightarrow \pi^+ \pi^- \eta(\eta^{(')})$ decay process.
The results show that the CP asymmetry of these four decay processes are significantly enhanced especially for the $\bar{B}^0_{s} \rightarrow \pi^+ \pi^- \bar K ^{0}$ decay process and the decay branching ratio also changes under resonance effect.
These work might provide support for the experimental analysis of the $\bar B^{0}_s$ meson.
\end{abstract}
\maketitle

\section{\label{intro}Introduction}
The non-leptonic decay of hadrons containing heavy quarks plays a crucial part in testing the Standard Model (SM) by examining the charge parity (CP) asymmetry mechanism in flavor physics \cite{N63,M73}. It can also gain our understanding of Quantum Chromodynamics and discovering new physical phenomenon beyond the SM.
The result of CP asymmetry relates to the weak phase in the Cabibbco-Kobayashi-Maskawa (CKM) matrix describing the mixing of quarks of different generations. Besides, a strong phase is also required to detect CP asymmetries \cite{T92}. Typically, this strong phase is provided by severals phenomenological models and the QCD loop corrections. Similarly, the generation of these phases may affect the decay process of vector mesons, CP asymmetries and even decay branching ratios. Nowadays, both theoretically and experimentally, there has been increased attention on CP asymmetry and decay branching ratio in the $B$ meson system.
The decay of $B$ meson has gradually changed from the analysis of two-body decay to the analysis of three-body decay, and has been widely carried out \cite{JJQ,SZ,15Li2022}.
In recent years, the BABAR, Belle, and LHCb collaborations have already measured a number of CP asymmetry and branching ratio parameters of three-body charmless $B$ decays in the experiment \cite{E01,E02,E03}. 
In the theory, a fully developed and widely used approach has been developed to calculate the hadron matrix elements of $B$ meson non-leptonic weak decay, including naive factorization \cite{M85,M87}, QCD factorization (QCDF) \cite{N1999,N2007,H2008}, Perturbative QCD (PQCD) \cite{Y01,YY01,CD01} approaches, soft collinear efficient theory (SCET) \cite{C01,C02} and factorization assisted topological-amplitude approach (FAT) \cite{HY20,XG21,XG22}.

Inspired by the achievements in two-body $B$ decays, we use a quasi-two-body approach to calculate the three-body decay process of the $B$ meson under the resonant effect in this work. The vector resonant effects are depicted by means of the common Breit-Wigner formalism, and a strong coupling is used to explain the subsequent two-body decay of the vector meson. Under the resonance contribution, vector meson dominance model (VMD) predicts that the vacuum polarisation of the photon is entirely made up of vector mesons of $\rho^{0}(770)$, $\omega(782)$ and $\phi(1020)$ \cite{NM67}. The transitions of $\omega(782)$ and $\phi(1020)$ decay to $\pi^+ \pi^-$ meson pairs which originate from isospin breaking related to the mixings of $\omega(782)-\rho^{0}(770)$ and $\phi(1020)-\rho^{0}(770)$. 
Since the decay rate of $ \rho^{0}(770) \rightarrow \pi^+ \pi^-$ is 100$\%$ \cite{PDG}, the mixing of intermediate particles $\rho^{0}(770)$, $\omega(782)-\rho^{0}(770)$ and $\phi(1020)-\rho^{0}(770)$ are mainly considered in our theory, ignoring the interference from other processes.
Besides, the strong phase from the three-body decay can be produced by intermediate resonance hadrons associated with the Breit-Wigner form. A new strong phase is formed under the mixing of the intermediate resonance hadrons, and combined with the weak phase calculation from the CKM matrix. We use a matrix composed of hadrons to combine the intermediate state of the decay process with the physical state of the isospin state \cite{2lu2022}. The interference caused by the three  mesons $\rho^{0}(770)$, $\omega(782)$ and $\phi(1020)$ can be solved by dynamical mechanisms. 
The analysis of vector meson resonance has greatly contributed to the understanding of particle properties and meson interactions \cite{MDA1979}.

In the previous calculation of CP asymmetry theory, we investigated the non-leptonic decay processes of $\bar{B}^0_{s} \rightarrow \rho^{0}(770)\pi^0(\bar K^{0}) \rightarrow \pi^+ \pi^- \pi^0(\bar K^{0})$ with the interference by PQCD \cite{P2017}. The PQCD approach combines QCD correction due to transverse momentum and introduces Sudakov factor to suppress the non-perturbative effect. Endpoint divergence is regulated by introducing the parton transverse momentum $k_T$ and the Sudakov factor at the expense of modeling the additional $k_T$ dependence of meson wave functions, and annihilation corrections are presented. The non-perturbative contribution is contained in the hadron wave function.
In the latest theoretical calculation of the decay branching ratio, in addition to the PQCD approach,
a framework of FAT approach is introduced \cite{SH23}, which includes the non-perturbation and non-factorization contributions of two-body $B$ decays. The decay amplitude of two-body charmless $B$ decay is divided into different electroweak topological Feynman figures under $SU(3)$ symmetry. By globally fitting all experimental data for these decays, the topological amplitudes including the nonfactorizable QCD contributions are extracted. However, the precision of this topological approach is limited by the size of the $SU(3)$ breaking effect.
In this paper, we use the QCDF approach to research and compare the results of the new calculation with the previous ones, and check the accuracy of the theoretical calculation of QCD. 
Within the framework of the QCDF, the analysis for the decay of $B$ meson can set the $b$-quark mass to infinity and ignore the higher-order contribution of $1/m_b$, the two-body non-lepton decay amplitude can be expressed as the producting of the form factor from the initial meson to the final meson and the light cone distribution amplitude of the final meson in the heavy quark limit. The logarithmically divergent integral is usually  parameterized in a model-independent manner and explicitly expressed as $\int_{0}^{1} dx/x\rightarrow \mathrm{X}_A$ \cite{2lu2021}. So we use the QCDF approach to study.
We calculate the CP asymmetry result of $\bar{B}^0_{s} \rightarrow V(\rho^{0}(770),\omega(782),\phi(1020))\pi^0(\bar K^{0}) \rightarrow \pi^+ \pi^- \pi^0(\bar K^{0})$ decay process and compare the influence of PQCD and QCDF on CP asymmetry under resonance effect. At the same time, the CP asymmetry results and local integration results of the two attenuation processes of $\bar{B}^0_{s} \rightarrow V(\rho^{0}(770),\omega(782),\phi(1020))\eta(\eta^{'}) \rightarrow \pi^+ \pi^- \eta(\eta^{'})$ are added.
Then we also calculate the decay branching ratio of these four decay processes under the resonance effect and the decay branching ratio without the resonance effect, and compare these results with the latest theoretical results. Since the PQCD approach only has the results of the direct decay process, but the branching ratios of the three-body $B$ decay studied under the FAT approach takes into account the virtual effects of the intermediate resonances $\rho^{0}(770), \omega(782), \phi(1020)$ on quasi-two-body decays, so we compare and discuss these results.
We will explore the role and contribution of resonance effects on the decay process $\bar B^{0}_s \rightarrow \pi^{+}\pi^{-}P$.

The overall structure of this paper is as follows.
In Sect. II, we introduce the resonance mechanism in subsect. A, briefly explain the QCDF approach in subsect. B, and show the amplitude invloving $\rho^{0}(770)$, $\omega(782)$ and $\phi(1020)$ interference in subsect. C. Then, in Sect. III, it mainly consists of the computational form of CP asymmetry, local integral form of CP asymmetry and the branching ratios of the three-body decay process, in subsect. A, B and C respectively. 
In Sect. IV, we analyze the curve results of CP asymmetry in these decay processes, calculate the local integral CP asymmetry and decay branching ratios in different phase space regions. The summary shows in Sect. V.

\section{\label{sum}Calculation of amplitude under QCDF}
\subsection{\label{subsec:form} The introduction of resonance mechanism}
Due to the vector meson dominance model (VMD), $e^{+}e^{-}$ can annihilate into a photon $\gamma$, the vacuum polarisation of the photon $\gamma$ which is dressed by coupling vector mesons $\rho^{0}(770)$, $\omega(782)$ and $\phi(1020)$, these vector mesons can then decay into $\pi^{+}\pi^{-}$ meson pairs and the VMD successfully describes the interaction between photons and hadron \cite{NM67}. 
The $\rho^{0}(770)-\omega(782)-\phi(1020)$ interference is caused by the difference in quark mass and electromagnetic interaction effects, they can decay directly into $\pi^{+}\pi^{-}$ meson pairs.
Besides, the transitions of $\omega(782)$ and $\phi(1020)$ decay to $\pi^{+}\pi^{-}$ meson pairs which originate in isospin breaking related to the mixing of $\omega(782)-\rho^{0}(770)$ and $\phi(1020)-\rho^{0}(770)$ \cite{2lu2022}.
We establish a resonance effect by considering the interference effects brought about by the mixing of three intermediate state particles($\rho^{0}(770)$, $\omega(782)$, $\phi(1020)$). Since the resonance effect is not a physical respresentation, a matrix is constructed to transform the isospin field into a physical representation. The relationship between the isospin field ($\rho_I, \omega_I, \phi_I$) and the physical respresentation ($\rho, \omega, \phi$) can be connected by the matrix, while ignoring the contribution of higher order terms. In order to easy to read, we use $\rho, \omega$ and $\phi$ to represent $\rho^{0}(770), \omega(782)$ and $\phi(1020)$. It can be expressed as:
\begin{equation}
	\left (
	\begin{array}{l}
		\rho  
		\\[0.5cm]
		\omega 
		\\[0.5cm]
		\phi 
	\end{array}
	\right )  =
	\left (
	\begin{array}{lll}
		\left<\rho_{I}|\rho \right> & \hspace{0.5cm} \left<\omega_{I}|\rho\right> & \hspace{0.5cm} \left<\phi_{I}|\rho\right> \\[0.5cm]
		\left<\rho_{I}|\omega\right> &  \hspace{0.5cm}\left<\omega_{I}|\omega\right> &  \hspace{0.5cm}\left<\phi_{I}|\omega\right>
		\\[0.5cm]
		\left<\rho_{I}|\phi\right> &  \hspace{0.5cm}\left<\omega_{I}|\phi\right> &  \hspace{0.5cm}\left<\phi_{I}|\phi\right>
	\end{array}
	\right )
	\left (
	\begin{array}{l}
		\rho_I  
		\\[0.5cm]
		\omega_I
		\\[0.5cm]
		\phi_I 
	\end{array}
	\right ) \\
	=
	\left (
	\begin{array}{lll}
		~~~~1 &\hspace{0.5cm} -F_{\rho\omega}(s) & \hspace{0.5cm}-F_{\rho\phi}(s) 
		\\[0.5cm]
		\displaystyle  F_{\rho\omega}(s) &  \hspace{0.5cm}~~~ 1 & \hspace{0.5cm}-F_{\omega\phi}(s)
		\\[0.5cm]
		\displaystyle  F_{\rho\phi}(s) &   \hspace{0.5cm} F_{\omega\phi}(s) &\hspace{0.5cm} ~~~1 
	\end{array}
	\right ),
	\label{L2}
\end{equation}
where $F_{VV}(s)$(V=$\rho, \omega, \phi$) is order $\mathcal{O}(\lambda)$, $(\lambda\ll 1)$ \cite{2lu2022}. Based on the isospin field $\rho_{I}(\omega_{I},\phi_{I})$, we can construct the isospin basis vectors $\left|I,I_3 \right>$, where $I$ and $I_3$ are isospin and its third components, respectively. Thus, physical states can be represented as linear combinations of basis vectors in a matrix.
We used orthogonal normalization to obtain the relationship between the physical state of the particle and the isospin basis vector,
so the physical manifestation of this form can be clearly expressed as $\rho=\rho_{I}-F_{\rho\omega}(s)\omega_{I}-F_{\rho\phi}(s)\Phi_{I}$,
$\omega=F_{\rho\omega}(s)\rho_{I}+\omega_{I}-F_{\omega\phi}(s)\Phi_{I}$,
$\Phi=F_{\rho\phi}(s)\rho_{I}+F_{\omega\phi}(s)\omega_{I}+\Phi_{I}$. 

Considering the physical and isospin representations, we make propagator definitions as $D_{V_1V_2}=\left< 0|TV_1V_2|0 \right> $ and $D_{V_1V_2}^{I}=\left< 0|TV_{1}^{I}V_{2}^{I}|0 \right>$, respectively. $V_{1}$ and $V_{2}$ of $D_{V_{1}V_{2}}$ refer to any two of the three particles $\rho$, $\omega$ and $\phi$ in physical. Bringing $\rho, \omega, \phi$ of the physical fields into the definition of $D_{V_1V_2}$, we find that the forms $D_{\rho\omega}$, $D_{\rho\phi}$ and $D_{\omega\phi}$ are identical. Due to there is no three vector meson mixing under the physical representation, $D_{V_{1}V_{2}}$ is equal to zero. We deduced these process about propagators in detail in our previous theory \cite{2lu2021}. In addition, according to the physical state expression of the two-vector meson mixing, the parameters of $F_{\rho\omega}$ is the order of $\mathcal{O}(\lambda)$ ($\lambda\ll 1$). Since the multiplication of multiple terms in the equation is a higher-order term, the result of the higher-order term is depressed, which can be ignored by us.
This paper only consider the process containing $\rho$ because the CP asymmetry of $\pi^{+}\pi^{-}$ meson pairs produced by the mixing process of $\omega$ and $\phi$ is almost absent under the resonance effect. We can define new mixing parameters based on decay width and mass, the specific expression is \cite{luG2024}
\begin{equation}
\begin{array}{l}
 \Pi_{\rho\omega}=F_{\rho\omega}(s-m^{2}_{\rho}+im_{\rho}\Gamma_{\rho})-F_{\rho\omega}(s-m^{2}_{\omega}+im_{\omega}\Gamma_{\omega}),
\\[0.2cm]
 \Pi_{\rho\phi}=F_{\rho\phi}(s-m^{2}_{\rho}+im_{\rho}\Gamma_{\rho})-F_{\rho\phi}(s-m^{2}_{\phi}+im_{\phi}\Gamma_{\phi}),
\end{array}  
\label{A}
\end{equation}
where $\Gamma_{V}$ and $m_{V}$ represents the decay width and mass of vector mesons $V$ $(V= \rho, \omega, \phi)$, respectively. 
The propagator $s_{V}$ of vector meson is associated with the invariant mass $\sqrt s$, which can be represented as 
$s_V=s-m_{V}^2+im_{V} \Gamma_{V}$. 
The vector resonance $\Gamma_{V}(s)$ for the energy-dependent width can be written \cite{JM47}:
\begin{eqnarray}
\Gamma_{V}(s)=\Gamma_{0}(\frac{q}{q_0})^{3}(\frac{m_V}{\sqrt{s}})X^{2}(qr_{BW}),
\label{NWA}
\end{eqnarray} 
where the expression for the Blatt-Weisskopf barrier factor $X^{2}(qr_{BW})$ is $\sqrt{\left[1+(q_{0}r_{BW})^2\right]/\left[1+(qr_{BW})^2\right]}$  , $q= \frac{1}{2} \sqrt{\left[s - (m_{\pi^+} + m_{\pi^-})^2\right]\left[s - (m_{\pi^+} - m_{\pi^-})^2\right]/s}$ is the momentum of the final state $\pi^+$ or $\pi^-$ in the rest frame of the resonance $V$, and $q_0$ is the value of $q$ when $s=m^2_V$. The value of the barrier radius
$r_{BW}$ is $4.0(GeV)^{-1}$ for all resonances \cite{RA46}. We consider the vector resonance state for the full width value $\Gamma_{0}$, which come from PDG of the decay fraction of $\rho$ to $\pi^{+}\pi^{-}$ is $100\%$ with the full width of $\rho$ being 149.1$\pm$0.8 MeV, $\omega$ to $\pi^{+}\pi^{-}$ is $1.53^{+0.11}_{-0.13}\%$ with the full width of $\omega$ being 8.68$\pm$0.13 MeV and $\phi$ to $\pi^{+}\pi^{-}$ is $(7.3\pm 1.3)\times10^{-5}\%$ with the full width of $\phi$ being 4.249$\pm$0.013 MeV in this paper \cite{PDG,J2003}. These data will also be used in the subsequent calculation of the decay width.

Besides, the mixing parameters of $\omega-\rho$ and $\phi-\rho$ are extracted from the $e^{+}e^{-}\rightarrow \pi^{+}\pi^{-}$ experimental data \cite{MN2000,P2009}.
To better interpret the mixing of $\omega-\rho$ and $\phi-\rho$, we define 
\begin{eqnarray}
	\widetilde{\Pi}_{\rho\omega}=\frac{(s-m^{2}_{\rho}+im_{\rho}\Gamma_{\rho})\Pi_{\rho\omega}}
	{(s-m^{2}_{\rho}+im_{\rho}\Gamma_{\rho})-(s-m^{2}_{\omega}+im_{\omega}\Gamma_{\omega})},
	\label{A} \quad \quad
	\widetilde{\Pi}_{\rho\phi}=\frac{(s-m^{2}_{\rho}+im_{\rho}\Gamma_{\rho})\Pi_{\rho\phi}}
	{(s-m^{2}_{\rho}+im_{\rho}\Gamma_{\rho})-(s-m^{2}_{\phi}+im_{\phi}\Gamma_{\phi})}.
	\label{A}
\end{eqnarray}

The mixing parameter of $\widetilde{\Pi}_{\rho\omega}(s)$ and $\widetilde{\Pi}_{\rho\phi}(s)$ are the momentum dependence of $\omega-\rho$ and $\phi-\rho$ interference, respectively \cite{P2011}. These parameters are functions of the momentum. Wolfe and Maltman have measured these parameters \cite{P2009}. $\widetilde{\Pi}_{\rho\omega}(s)$ and $	\widetilde{\Pi}_{\rho\phi}(s)$ can be expressed in the form of real and imaginary parts:
\begin{eqnarray}
	\widetilde{\Pi}_{\rho\omega}(s)={\mathfrak{Re}}\widetilde{\Pi}_{\rho\omega}(m_{\omega}^2)+{\mathfrak{Im}}\widetilde{\Pi}_{\rho\omega}(m_{\omega}^2),
	\label{A} \quad \quad
	\widetilde{\Pi}_{\rho\phi}(s)={\mathfrak{Re}}\widetilde{\Pi}_{\rho\phi}(m_{\phi}^2)+{\mathfrak{Im}}\widetilde{\Pi}_{\rho\phi}(m_{\phi}^2).
	\label{A}
\end{eqnarray}
Numerical results for the real and imaginary parts of the $\omega-\rho$($\phi-\rho$) mixing parameter $\widetilde\Pi_{\rho\omega}$($\widetilde\Pi_{\rho\phi}$) at $s=m^2_{\omega}$($s=m^2_{\phi}$) can be given \cite{P2011}:
\begin{equation}
\begin{array}{l}
\mathfrak{Re}\widetilde{\Pi}_{\rho\omega}(m_{\omega}^2)=-4760\pm440
\rm{MeV}^2,\quad
{\mathfrak{Im}}\widetilde{\Pi}_{\rho\omega}(m_{\omega}^2)=-6180\pm3300
\textrm{MeV}^2;  
\\[0.2cm]
\mathfrak{Re}\widetilde{\Pi}_{\rho\phi}(m_{\phi}^2)=796\pm312
\rm{MeV}^2,\quad 
{\mathfrak{Im}}\widetilde{\Pi}_{\rho\phi}(m_{\phi}^2)=-101\pm67
\textrm{MeV}^2.
\end{array}  
\label{A}
\end{equation}

\subsection{\label{subsec:form}The QCD factorization}
M.Beneke et al. believe that the form factor of the $B$ to the final hadron transition is mainly contributed from the non-perturbed region, the non-factorization effect of hadron matrix elements is mainly the exchange of hard gluons in the two-body non-light decay of $B$ meson. They propose a new approach for computing hadron matrix elements which is QCD factorization \cite{N1999,N2003}.
The low energy effective Hamiltonian form for the non-light and weak decay of $B$ meson decays can be written as \cite{G1996,JF2002}:
\begin{eqnarray}
\mathcal{H}_{eff}=\frac{G_F}{\sqrt2} \sum_{q =u,c} V_q  \left\{C_{1}(\mu)Q^{q}_{1}(\mu)+C_{2}(\mu)Q^{q}_{2}(\mu)+\sum^{10}_{k=3}C_{K}(\mu)Q_{K}(\mu)+C_{7\gamma}(\mu)Q_{7\gamma}(\mu)+C_{8g}(\mu)Q_{8g}(\mu)\right\}+H.c.,
\label{A}
\end{eqnarray}
where $V_q$ is the factor associated with the CKM matrix element. The Wilson parameter $C_i$ can be calculated using perturbation theory and renormalization group approaches. $Q_i$ is a valid operator for localization \cite{JF2002}: 

the current-current operator which are
\begin{equation}
\begin{array}{l}
Q^{u}_{1}=(\bar u_{\alpha} b_{\alpha})_{V-A}(\bar q_{\beta} u_{\beta})_{V-A},\quad
Q^{c}_{1}=(\bar c_{\alpha} b_{\alpha})_{V-A}(\bar q_{\beta} c_{\beta})_{V-A}, 
\\[0.2cm]
Q^{u}_{2}=(\bar u_{\alpha} b_{\beta})_{V-A}(\bar q_{\beta} u_{\alpha})_{V-A},\quad 
Q^{c}_{2}=(\bar c_{\alpha} b_{\beta})_{V-A}(\bar q_{\beta} c_{\alpha})_{V-A},
\end{array}  
\label{A}
\end{equation}

the QCD penguins operator which are
\begin{equation}
\begin{array}{l}
Q_{3}=(\bar q_{\alpha} b_{\alpha})_{V-A}\sum_{q^{'}}(\bar q_{\beta}^{'} q_{\beta}^{'})_{V-A},\quad
Q_{4}=(\bar q_{\beta} b_{\alpha})_{V-A}\sum_{q^{'}}(\bar q_{\alpha}^{'} q_{\beta}^{'})_{V-A},
\\[0.2cm]
Q_{5}=(\bar q_{\alpha} b_{\alpha})_{V-A}\sum_{q^{'}}(\bar q_{\beta}^{'} q_{\beta}^{'})_{V+A},\quad 
Q_{6}=(\bar q_{\beta} b_{\alpha})_{V-A}\sum_{q^{'}}(\bar q_{\alpha}^{'} q_{\beta}^{'})_{V+A},
\end{array}  
\label{A}
\end{equation}

the electroweak penguins operator which are
\begin{equation}
\begin{array}{l}
Q_{7}=\frac{3}{2}(\bar q_{\alpha} b_{\alpha})_{V-A}\sum_{q^{'}}e_{q^{'}}(\bar q_{\beta}^{'} q_{\beta}^{'})_{V+A},\quad
Q_{8}=\frac{3}{2}(\bar q_{\beta} b_{\alpha})_{V-A}\sum_{q^{'}}e_{q^{'}}(\bar q_{\alpha}^{'} q_{\beta}^{'})_{V+A},
\\[0.2cm]
Q_{9}=\frac{3}{2}(\bar q_{\alpha} b_{\alpha})_{V-A}\sum_{q^{'}}e_{q^{'}}(\bar q_{\beta}^{'} q_{\beta}^{'})_{V-A},\quad
Q_{10}=\frac{3}{2}(\bar q_{\beta} b_{\alpha})_{V-A}\sum_{q^{'}}e_{q^{'}}(\bar q_{\alpha}^{'} q_{\beta}^{'})_{V-A},
\end{array}  
\label{A}
\end{equation}

and the magnetic penguins operator which are
\begin{equation}
\begin{array}{l}
Q_{7\gamma}=\frac{e}{8\pi^{2}}m_b \bar q_{\alpha} \sigma^{\mu v}(1+\gamma_5)b_{\alpha}F_{\mu v},\quad
Q_{8g}=\frac{g}{8\pi^{2}}m_b \bar q_{\alpha} \sigma^{\mu v}(1+\gamma_5)t^{a}_{\alpha\beta}b_{\beta}G^{a}_{\mu v},
\end{array}  
\label{A}
\end{equation}
where $(\bar q_{i}q_{j})_{V\pm A}\equiv \bar q_{i}\gamma_{\mu}(1\pm\gamma_5)q_{j}$, $q^{'}$ corresponds to a certain energy scale, and we can take the flavor of all the free quarks at this energy scale. The weak decay of the $B$ meson is $\mu \sim O(m_b) $ in the energy scale and $q^{'}\in \left\{u,d,s,c,b\right\}$; $e_{q^{'}}$ is the charge of the quark $q^{'}$; $\alpha$ and $\beta$ are color indices.

When dealing with the decay of the baryons into two mesons $M_1$ and $M_2$, the decay amplitudes are usually divided into emission and annihilation parts according to the topology structure. At the heavy quark limit, the emission part can be expressed as the product of the decay constant and the form factor, while the weak annihilation part is generally considered to be suppressed by power. 
According to Eq. (7), the decay amplitude of the $B$ meson to the final state of the emission part($\mathcal{A}_E$) and the weak annihilation part($\mathcal{A}_W$) have the following form \cite{JF20031}:
\begin{eqnarray}
\mathcal{A}_E(B\rightarrow M_1 M_2)=\frac{G_F}{\sqrt2} \sum_{q =u,c}\sum_{i} V_q  \alpha_{i}^{q}(\mu)\left<M_1 M_2\left|Q_{i}\right|B\right>_{F},
\\
\mathcal{A}_W(B\rightarrow M_1 M_2)=\frac{G_F}{\sqrt2} \sum_{q =u,c}\sum_{i} V_q  f_B f_{M_1} f_{M_2} b_{i}(M_1,M_2),
\label{A}
\end{eqnarray}
where $\alpha_{i}^{q}(\mu)$ are flavour parameters which can be expressed in terms of the effective parameters $\alpha_{i}^{q}$, which can be calculated perturbatively and has been demonstrated \cite{N2003}.
$f_B$, $f_{M_1}$ and $f_{M_2}$ are the decay constant associated with the initial and final meson, the numerical results are usually extracted using experimental approaches. 

In the heavy quark limit, ignoring the correction of $\Lambda_{QCD}/m_{b}$ to the leading order, the effective operator of hadron matrix element $\left<M_1 M_2\left|Q_{i}\right|B\right>$ can be calculated by the following formula by using the QCDF:
\begin{eqnarray}
\begin{array}{l}
\left<M_1 M_2\left|Q_{i}\right|B\right>=\sum_{j} F_{j}^{B\rightarrow M_1}\int_{0}^{1}d{x} T^{I}_{ij}(x)\Phi_{M_2}(x)+(M_1\leftrightarrow M_2)
\\[0.2cm]
\quad\quad\quad\quad\quad\quad \quad+\int_{0}^{1}d{\xi}\int_{0}^{1}dx\int_{0}^{1}dyT^{II}_{i}(\xi,x,y)\Phi_{B}(\xi)\Phi_{M_1}(x)\Phi_{M_2}(y),
\end{array}  
\label{A}
\end{eqnarray}
where $F_{j}^{B\rightarrow M_1}$ represents the transition form factor of $B\rightarrow M_1$, $T^{I}_{ij}$ and $T^{II}_{i}$ are the computable hard scattering part of perturbation theory, respectively, $\Phi_{X}(x)$ is the optical cone distribution amplitude of the quark——Fock state of the hadron, where the final hadrons $M_1$ and $M_2$ are both light mesons, or $M_1$ is a light meson and $M_2$ is a heavy quark even element. Then, when $M_1$ is a heavy meson and $M_2$ is a light meson, the hadron matrix element form is:
\begin{eqnarray}
\begin{array}{l}
\left<M_1 M_2\left|Q_{i}\right|B\right>=\sum_{j} F_{j}^{B\rightarrow M_1}\int_{0}^{1}d{x} T^{I}_{ij}(x)\Phi_{M_2}(x),
\end{array}  
\label{A}
\end{eqnarray}

The calculation of hadron matrix elements in the two-body decay of $B$ meson becomes more convenient by Eqs (14) and (15). 
The non-perturbation effect is manifested in the amplitude and shape factor of the meson optical cone distribution. The form factor $F_{j}^{B\rightarrow M_1}$ is a physical quantity that contains both hard and soft contributions (so the hard contribution needs to be subtracted from the hard scattering function $T^{I}_{ij}$ and $T^{II}_{i}$). 
This form factor can be determined from experiments on the semi-light decay of $B$ meson or QCD theory. The light cone distribution amplitude of mesons can also be extracted from other hard scattering processes. The leading order of the decay amplitude is the contribution of naive factorization. In the heavy quark limit, the radiation correction of the leading order can be calculated to all orders of $\alpha_s$ without thinking about the $1/m_b$ power correction.

Similarly, the weak annihilation contributions are described by the terms $b_{i}$ and $b_{i}^{EW}$. They are expressed in the following ways:
\begin{equation}
\begin{array}{l}
b_{1}(M_1,M_2)=\frac{C_F}{N^{2}_c}C_{1}A^{i}_{1}(M_1,M_2),\quad
b_{2}(M_1,M_2)=\frac{C_F}{N^{2}_c}C_{2}A^{i}_{1}(M_1,M_2), 
\\[0.2cm]
b_{3}(M_1,M_2)=\frac{C_F}{N^{2}_c}\left\{C_{3}A^{i}_{1}(M_1,M_2)+C_{5}A^{i}_{3}(M_1,M_2)+\left[C_{5}+N_{c}C_{6}\right]A^{f}_{3}(M_1,M_2)\right\},
\\[0.2cm]
b_{4}(M_1,M_2)=\frac{C_F}{N^{2}_c}\left\{C_{4}A^{i}_{1}(M_1,M_2)+C_{6}A^{i}_{2}(M_1,M_2)\right\},
\\[0.2cm]
b^{EW}_{3}(M_1,M_2)=\frac{C_F}{N^{2}_c}\left\{C_{9}A^{i}_{1}(M_1,M_2)+C_{7}A^{i}_{3}(M_1,M_2)+\left[C_{7}+N_{c}C_{8}\right]A^{f}_{3}(M_1,M_2)\right\},
\\[0.2cm]
b^{EW}_{4}(M_1,M_2)=\frac{C_F}{N^{2}_c}\left\{C_{10}A^{i}_{1}(M_1,M_2)+C_{8}A^{i}_{2}(M_1,M_2)\right\},
\end{array}  
\label{A}
\end{equation}
where the annihilation coefficients $b_{1,2}$ correspond to the current-current operator $Q_{1,2}$, the coefficients $b_{3,4}$ correspond to the QCD penguins operator $Q_{3\sim6}$, and the annihilation coefficients $b^{EW}_{3,4}$ correspond to the electroweak penguins operator $Q_{7\sim10}$. The amplitude $A^{i,f}_{n}(n = 1,2,3)$ comes from the annihilation contribution.
The superscript $i$ refers to gluon emission from the initial-state quark and the superscript $f$ refers to gluon emission from the final-state quark. The form of $A^{i,f}_{n}$ has been confirmed in Ref \cite{N2003}. 
Besides, we handle the endpoint integrals of these logarithmic divergences arising from the hard scattering process involving the spectator quark. We employ phenomenological parameters to express it in the form of $\mathrm{X}=\int_{0}^{1} dx/x $ by Ref \cite{N1999}. The QCDF has been shown to be effective in the non-light and weak decay of $B$ meson, including chirally enhanced corrections.

\subsection{\label{subsec:form} The amplitude of $\bar B^0_{s}\rightarrow \pi^{+}\pi^{-}P$ involved $\rho-\omega-\phi$ interference}
\begin{figure}[h]
	\centering
	\includegraphics[height=8cm,width=12cm]{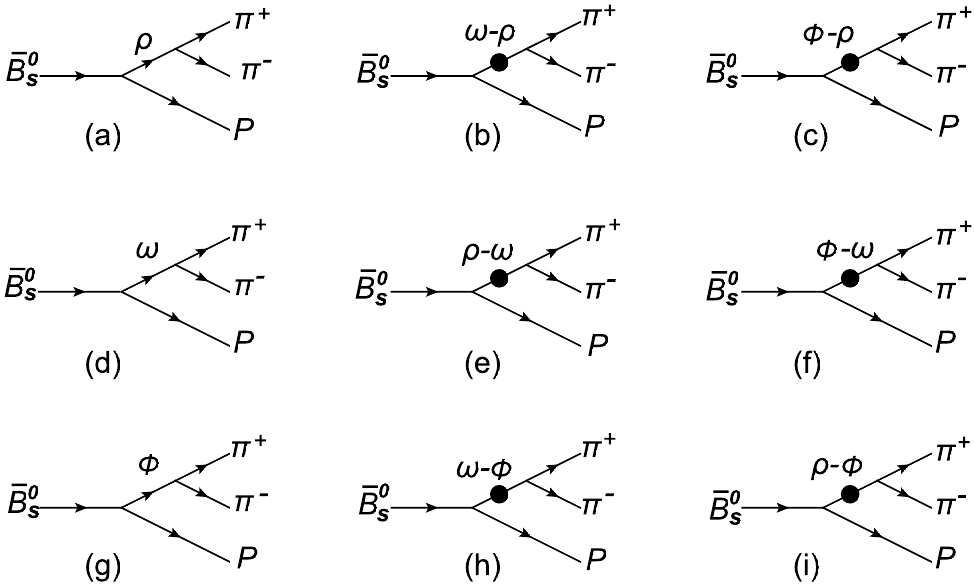}
	\caption{The decay processes for the channel of $\bar B^{0}_{s}\rightarrow \rho(\omega,\phi)P \rightarrow \pi^{+}\pi^{-}P$.}
	\label{fig1}
\end{figure}

Under the framework of QCDF, we analysis the effect of resonance effect generated by mixing of intermediate particles for CP asymmetry and branching ratio of $\bar B^0_{s}$. The decay processes $\bar B^0_{s}\rightarrow \pi^{+}\pi^{-}P$ involved $\rho(\omega,\phi)$ meson is shown in FIG. \ref{fig1}, where $P$ represents the final pseudo-scalar meson.
Considering the $\rho \rightarrow\pi^{+}\pi^{-}$ process from isospin symmetry breaking, we also believe that the vector meson to $\pi^{+}\pi^{-}$ can be attributed to the existence of $\omega-\rho$ and $\phi-\rho$ mesons intermediate state of resonance effect.
So we only consider the decay process of the key intermediate state particle $\rho \rightarrow\pi^{+}\pi^{-}$ and do not take the $\rho^{-} \rightarrow\pi^{-}\pi^{0}$ decay process and $\rho^{+} \rightarrow\pi^{+}\pi^{0}$ decay process into account in our approach.
The influence of resonance effect on CP asymmetry which considering the interference of intermediate particles are shown in the six figures (b), (c), (e), (f), (h) and (i) of FIG. \ref{fig1}. Taking FIG. (b) as an example, $\bar B^0_{s}$ decays into $\omega$ mesons and pseudo-scalar meson $P$, $\omega$ can decay into $\rho$ meson firstly and then $\rho$ meson decay into $\pi^{+}\pi^{-}$ mesons. In this process, the resonance effect generated by the interaction between $\omega$ and $\rho$ meson is involed, and the CP asymmetry and branching ratio in the resonance state are calculated. 

To the first leading order of isospin breaken, the most contributing is from (b) and (c) of the six plots.
Therefore, the CP asymmetry results of figures (a), (b) and (c) are calculated with emphasis. The CP asymmetry result is significantly depressed for the others processes which the decay rate is relatively low under the resonance effect, so we do not consider them. 

We employ a quasi-two-body decay process to calculate the CP asymmetry and branching ratio.   
In the two-body decay of the $B$ meson, the form factor governing the transition from initial hadron to the final hadron is dominated by non-perturbative effects \cite{P2022}.
We can calculate the perturbative contribution associated with the hard gluon from the QCD correction. In the three-body decay process, we will adopt the naive Breit-Wigner form for $\rho$ with the pole mass $m_{\rho}$ = 0.775 GeV and the width $\rho$ =0.149 GeV \cite{HYC42}. For example, the decay process of $\bar{B}^{0}_{s}\rightarrow \rho(\rho \rightarrow \pi^{+}\pi^{-})\pi^0$ can be expressed as: 
\begin{eqnarray}
\mathcal{M}_{\rho}=\frac{\left< \rho\pi^0\left|\mathcal{H}_{eff}\right|\bar B^{0}_{s}\right>\left<\pi^+\pi^-\left|\mathcal{H}_{\rho \pi^+\pi^-}\right|\rho\right>}{s-m^{2}_{\rho}+im_{\rho}\Gamma_{\rho}},
\label{NWA}
\end{eqnarray} 
where $\mathcal{H}_{\rho \pi^+\pi^-}$ is the effective Hamiltonians of the strong processes $\rho \rightarrow \pi^+ \pi^- $, $s-m^{2}_{\rho}+im_{\rho}\Gamma_{\rho}$ is the Breit-Wigner form for the propagator of $\rho$ and $s$ is the invariant mass squared of mesons $\pi^+$ and $\pi^-$.

We give the decay amplitude due to quasi-two-body decay with emission and annihilation contributions for $\bar{B}^{0}_{s}\rightarrow \rho(\rho \rightarrow \pi^{+}\pi^{-})\pi^0$ under QCDF.
The remaining amplitude forms of $B\rightarrow VP\rightarrow \pi^{+}\pi^{-} P$ are given in Appendix A:
\begin{eqnarray}
\begin{array}{c}
\mathcal{M}\left(\bar{B}^{0}_{s}\rightarrow \rho(\rho \rightarrow \pi^{+}\pi^{-})\pi^0 \right)=$$\sum_{q =u,c}$$ {\frac{G_F g_{\rho \pi^{+}\pi^{-} }f_{B_{s}}f_{\pi}f_{\rho}}{2\sqrt{2}s_{\rho }}}\left\{V_{ub}V_{us}^{*}\left[b_{1}(\pi,\rho)\right.\right.\\
\\
\left.\left.+b_{1}(\rho,\pi)\right]-V_{tb}V_{ts}^{*}\left[2b_{4}(\pi,\rho)+2b_{4}(\rho,\pi)+\frac{1}{2}b_{4}^{EW}(\pi,\rho)+\frac{1}{2}b_{4}^{EW}(\rho,\pi)\right]\right\},
\end{array}
\end{eqnarray}
\begin{eqnarray}
	\begin{array}{c}
		\mathcal{M}\left(\bar{B}^{0}_{s}\rightarrow \omega(\omega \rightarrow \pi^{+}\pi^{-})\pi^0 \right)=$$\sum_{q =u,c}$$ {\frac{G_F g_{\omega\pi^{+}\pi^{-} }f_{B_{s}}f_{\pi}f_{\omega}}{2\sqrt{2}s_{\omega}}} \\
		\\
		\left\{V_{ub}V_{us}^{*}\left[b_{1}(\pi,\omega)+b_{1}(\omega,\pi)\right]-V_{tb}V_{ts}^{*}\left[\frac{3}{2}b_{4}^{EW}(\pi,\omega)+\frac{3}{2}b_{4}^{EW}(\omega,\pi)\right]\right\},
	\end{array}
\end{eqnarray}
\begin{eqnarray}
\begin{array}{c}
\mathcal{M}\left(\bar{B}^{0}_{s}\rightarrow \phi(\phi \rightarrow \pi^{+}\pi^{-})\pi^0 \right) =$$\sum_{q =u,c}$${\frac{G_F g_{\phi  \pi^{+}\pi^{-} }m_{\phi}\epsilon(\lambda) \cdot p_{\pi}}{s_{\phi}}}  \\
\\
 \left\{V_{ub}V_{us}^{*} f_{\pi}A_{0}^{B^{0}_{s} \rightarrow \phi}a_2 +V_{tb}V_{ts}^{*}\left[f_{\pi}A_{0}^{B^{0}_{s} \rightarrow \phi}(\frac{3}{2}a_7-\frac{3}{2}a_9)\right]\right\},
\end{array}
\end{eqnarray}
where $V_{ub}V^{*}_{us}$ and $V_{tb}V^{*}_{ts}$ are CKM matrix elements, $s_{\rho}$, $s_{\omega}$ and $s_{\phi}$ are the Breit-Wigner factors. The decay constants  $f_{\pi}$, $f_{B_s}$ and $f_{\rho(\omega,\phi)}$ correspond to the non-perturbative contributions, while the coefficients $a_{1,2...n}$ are associated with the Wilson coefficient $C_i$ \cite{PRD2000}. The form factors for process $B\rightarrow\phi$ is denoted by $A_0^{B^{0}_{s}\rightarrow\phi}$ which arise from non-perturbative effects. Additionally, annihilation contributions given by $b_1$, $b_4$ and $b_4^{EW}$ have been considered in Ref.\cite{J2016}. Interestingly, the contribution of the $\bar{B}^{0}_{s}\rightarrow \rho(\omega) \pi^0 \rightarrow \pi^{+}\pi^{-}\pi^0$ decay process is dominated by the annihilation contribution, but there is no annihilation contribution in the amplitude of the $\bar{B}^{0}_{s}\rightarrow \phi \pi^0 \rightarrow \pi^{+}\pi^{-}\pi^0$ decay process.
Here $\epsilon$ denotes the polarization vector meson and ${p_\pi}$ represents momentum of $\pi$. 
$g_{V \pi^{+}\pi^{-} }$ is the effective coupling constants which can be expressed in terms of the decay width of $V \rightarrow \pi^+\pi^-$. Furthermore, we define a mass parameter which characterizes the quark mass related to the meson component in $O_i$. The values of some input parameters and constants are given in Appendix B \cite{PDG,PR2000}.

\section{\label{CP}CP asymmetry and Branching ratio}
\subsection{\label{subsec:CP1}The form of strong phase of CP asymmetry}
The total amplitude used in the calculation is denoted by $\mathcal{M}$, where $\mathcal{M}$ represents the sum of the tree contribution ($\big<\pi^{+}\pi^{-}P|H^T|\bar B^{0}_{s}\big>$) and the penguin contribution ($\big<\pi^{+}\pi^{-}P|H^P|\bar B^{0}_{s}\big>$). 
The relative strong phase angle $\delta$ and weak phase angle $\phi$ affecting CP asymmetry, where the strong phase angle $\delta$ comes from the strong phase under the resonance effect and the weak phase angle $\phi$ comes from the CKM matrix.
After that, we define the new total amplitude by the ratio of the penguin contribution to the tree contribution. The expression can be defined \cite{2lu2021}:
\begin{eqnarray}
	\mathcal{M}=\big<\pi^{+}\pi^{-}P|H^T|\bar B^{0}_{s}\big>[1+re^{i(\delta+\phi)}],
	\label{A'}
\end{eqnarray}
where the parameter $r$ represents the ratio between the amplitude contributions of the penguin level and the tree level. Furthermore, considering the interference of $\omega-\rho$ and $\phi-\rho$, we can present the detailed form of the tree and penguin amplitude by combining aforementioned decay diagrams in FIG.\ref{fig1}:
\begin{eqnarray}
	\begin{array}{c}
		\big<\pi^{+}\pi^{-}P|H^T|\bar B^{0}_{s}\big>
		=\frac{g_{\rho \rightarrow \pi^+ \pi^- }T_{\rho}}{s-m^{2}_{\rho}+im_{\rho}\Gamma_{\rho}}
		+\frac{g_{\rho \rightarrow \pi^+ \pi^- }\widetilde{\Pi}_{\rho\omega}T_{\omega}}{(s-m^{2}_{\rho}+im_{\rho}\Gamma_{\rho})(s-m^{2}_{\omega}+im_{\omega}\Gamma_{\omega}) }
		+\frac{g_{\rho \rightarrow \pi^+ \pi^- }\widetilde{\Pi}_{\rho\phi}T_{\phi}}{(s-m^{2}_{\rho}+im_{\rho}\Gamma_{\rho})(s-m^{2}_{\phi}+im_{\phi}\Gamma_{\phi})},
	\end{array}
\end{eqnarray}
\begin{eqnarray}
	\begin{array}{c}
		\big<\pi^{+}\pi^{-}P|H^P|\bar B^{0}_{s}\big>=\frac{g_{\rho \rightarrow \pi^+ \pi^- }P_{\rho}}{s-m^{2}_{\rho}+im_{\rho}\Gamma_{\rho}}
		+\frac{g_{\rho \rightarrow \pi^+ \pi^- }\widetilde{\Pi}_{\rho\omega}P_{\omega}}{(s-m^{2}_{\rho}+im_{\rho}\Gamma_{\rho})(s-m^{2}_{\omega}+im_{\omega}\Gamma_{\omega}) }
		+\frac{g_{\rho \rightarrow \pi^+ \pi^- }\widetilde{\Pi}_{\rho\phi}P_{\phi}}{(s-m^{2}_{\rho}+im_{\rho}\Gamma_{\rho})(s-m^{2}_{\phi}+im_{\phi}\Gamma_{\phi})}
	\end{array}
\end{eqnarray}
where $T_{\rho(\omega,\phi)}$ and $P_{\rho(\omega,\phi)}$ are the Tree and penguin contribution for $\bar B^{0}_{s}\rightarrow\rho (\omega,\phi)P$ decay process, respectively. In this paper, the Tree contribution is associated with $V_{ub}V^{*}_{us}$ and the penguin contribution is associated with $V_{tb}V^{*}_{ts}$ in Eqs  (18)-(20) and Appendix A.
$s$ is the invariant mass squared of mesons $\pi^+\pi^-$ \cite{S2009}. 
Subsequently, by using the Tree and penguin contributions, we are able to derive new strong phases. These include: $ P_{\omega}/T_{\rho} \equiv r_{1}e^{i(\delta_\lambda+\phi)}$, $P_{\phi}/T_{\rho} \equiv r_{2}e^{i(\delta_x+\phi)}$, $T_{\omega}/T_{\rho} \equiv r_{3} e^{i\delta_\alpha}$, $T_{\phi}/T_{\rho} \equiv r_{4} e^{i\delta_\tau}$ and $P_{\rho}/P_{\omega} \equiv r_{5} e^{i\delta_\beta}$. Here $\delta_\lambda$, $\delta_x$, $\delta_\alpha$, $\delta_\tau$ and $\delta_\beta$  represent strong phases.
The obtained results can be substituted into Eq. (21). After simplification, we can obtain 
\begin{eqnarray}
	\begin{split}
		re^{i\delta}=
		\frac{r_{1}e^{i\delta_\lambda}r_{5} e^{i\delta_\beta}s_\phi s_{\omega}+r_{2}
			e^{i\delta_x}s_{\omega}\widetilde{\Pi}_{\rho\phi}+r_{1}e^{i\delta_\lambda}s_\phi\widetilde{\Pi}_{\rho\omega}}
		{r_{4} e^{i\delta_\tau}s_{\omega}\widetilde{\Pi}_{\rho\phi}+s_\phi s_{\omega}+r_{3}e^{i\delta_\alpha}s_\phi\widetilde{\Pi}_{\rho\omega}}.
	\end{split}
\end{eqnarray}
The weak phase $\phi$ is determined by the ratio of $V_{ub}V^{*}_{ud}$ to $V_{tb}V^{*}_{td}$ or the ratio of $V_{ub}V_{us}^{*}$ to $V_{tb}V_{ts}^{*}$ in the CKM matrix. We conclude that ${\rm sin}\phi =  \eta/\sqrt{(\rho-\rho^2-\eta^2)^2+\eta^2}$ and ${\rm cos}\phi = (\rho-\rho^2-\eta^2)/\sqrt{(\rho-\rho^2-\eta^2)^2+\eta^2}$, or ${\rm sin}\phi = - \eta/\sqrt{\rho^2+\eta^2}$ and ${\rm cos}\phi = - \rho/\sqrt{\rho^2+\eta^2}$ by Wolfenstein parameters \cite{MN2000}. 

\subsection{\label{subsec:form}Local integral form of CP asymmetry}
We will provide a reference for future experiments by integrating $A_{CP}$ over the phase space in this subsection. The amplitude for the decay process of $\bar{B}_s^{0} \rightarrow \rho \pi^0$ can be given by $M_{\bar B_s^{0}\rightarrow \rho \pi^0}^{\lambda}$=$\alpha p_{\bar B_s^{0}} \cdot \epsilon^{*}(\lambda)$,
where $\mathit{\lambda}$ is the direction of polarization for $\mathit{\epsilon}$.  $\mathit{\epsilon}$ is the $\rho$ mean polarization vector. $p_{\bar B_s^{0}}$ is the $\bar{B}_s^{0}$ meson's momentum. $\mathit{\alpha}$ represents the part of the amplitude which is independent of $\mathit{\lambda}$. The decay process $\rho\rightarrow \pi^+ \pi^-$ can be shown as $M_{\rho \rightarrow \pi^{+} \pi^{-}}^{\lambda}=g_{\rho}\epsilon(\lambda)\left(p_{1}-p_{2}\right)$, where $p_{1}$ and $p_{2}$ denote the momenta of $\pi^{+}$ and $\pi^{-}$ generated by the $\rho$ meson \cite{EP2013,J2020}.
So, the total amplitude of $\bar B_s^{0}\rightarrow\rho \pi^0 \rightarrow \pi^{+}\pi^{-} \pi^{0}$ decay process can be known as 
\begin{equation}
\begin{aligned}
A =\alpha p_{\bar B_s^{0}}^{\mu} \frac{\sum_{\lambda} \epsilon_{\mu}^{*}(\lambda) \epsilon_{\nu}(\lambda)}{s_{\rho}} \frac{g_{\rho}}{s_{\rho}}\left(p_{1}-p_{2}\right)^{\nu},
\end{aligned}
\end{equation}
where $\sqrt{s}$ and $\sqrt{s^{\prime}}$ represent the low and high invariant mass of the $\pi^{+}\pi^{-}$ pair, $s_{\max }^{\prime}$ and $s_{\min }^{\prime}$ are the maximum and  minimum values of $s^{\prime}$ for a fixed $s$, respectively \cite{R2014}. We get $m^{2}_{ij}$=$p^{2}_{ij}$ by conservation of momentum and energy during the three-body decay process. Therefore, the amplitude can be written as follow:
\begin{equation}
A =\frac{g_{\rho}}{s_{\rho}} \cdot \frac{M_{\bar B_s^{0}\rightarrow \rho \pi^{0}}^{\lambda}}{p_{\bar B_s^{0}} \cdot \epsilon^{*}} \cdot\left(\sigma-s^{\prime}\right)=\left(\sigma-s^{\prime}\right)\cdot \mathcal{M},
\end{equation}
where $\mathcal{M}$ is the substitution of the previous formula, $\sigma$ is taken to be a constant wich is related to $s$ and it can be see as $\sigma=\frac{1}{2}\left(s_{\max }^{\prime}+s_{\min }^{\prime}\right)$. For a fixed $s$, the differential CP asymmetry parameter can
be defined as $A_{CP}=( |\mathcal{M}|^{2}-|\overline{\mathcal{M}}|^{2})/(|\mathcal{M}|^{2}+|\overline{\mathcal{M}}|^{2})$. Then we integrate the denominator and numerator of $A_{CP}$ within the range of $\Omega \left(s_{1}<s<s_{2}, s_{1}^{\prime}<s^{\prime}< s_{2}^{\prime}\right)$. The localized integrated CP asymmetry that shows in this form \cite{WZWG2015}:
\begin{equation}
A_{CP}^{\Omega}=\frac{\int_{s_{1}}^{s_{2}} \mathrm{~d} s \int_{s_{1}^{\prime}}^{s_{2}^{\prime}} \mathrm{d} s^{\prime}\left(\sigma-s^{\prime}\right)^{2}\left(|\mathcal{M}|^{2}-|\overline{\mathcal{M}}|^{2}\right)}{\int_{s_{1}}^{s_{2}} \mathrm{~d} s \int_{s_{1}^{\prime}}^{s_{2}^{\prime}} \mathrm{d} s^{\prime}\left(\sigma-s^{\prime}\right)^{2}\left(|\mathcal{M}|^{2}+|\overline{\mathcal{M}}|^{2}\right)}.
\end{equation}
 Due to $s$ varies in a small region, $\sigma$ can be treated as a constant approximately. Thus, we can cancel the influence of $\int_{s_{1}^{'}}^{s_{2}^{'}}{ds^{'}}\left( \sigma -s^{'} \right) ^2$ \cite{WC2017}. 
It is assumed that $s_{\min}^{'}<s^{'}<s_{\max}^{'}$ represents an integral interval of the high invariance mass of $\pi^+ \pi^-$, while $\int_{s_{\min}^{'}}^{s_{\max}^{'}}{ds^{'}}\left( \sigma -s^{'} \right) ^2$ represents the factor that is dependent upon $s$. 

\subsection{\label{subsec:form}Decay branching ratio under resonance effect}
Due to isospin breaking, the effects of three-particle mixing on the branching ratios of $\bar{B}_{s}^{0}\rightarrow \rho (\omega,\phi)\pi^{0}(\bar K^{0}, \eta,\eta^{'})\rightarrow \pi^{+}\pi^{-} \pi^{0}(\bar K^{0}, \eta,\eta^{'})$ are symmetrical. By considering the value of $g_{\rho \rightarrow \pi^+ \pi^-}$ \cite{SH23}, we calculate the branching ratios of $ \bar{B}_{s}^{0} \to \pi^{+}\pi^{-} \pi^{0}(\bar K^{0}, \eta,\eta^{'}) $ under the three-particle mixing. 
The formula for the differential branching ratios originates from the S-wave contribution by PDG \cite{PDG}. But we consider the process that primarily involves the P-wave contribution \cite{BA51}. 
The differential branching ratios for the quasi-two-body $ \bar{B}_{s}^{0} \to  VP \to  \pi^{+}\pi^{-} P $ decays is written as \cite{JC21,OL,BA51}:  
\begin{equation}  
\frac{dB}{d\xi} = \frac{\tau_{B_{s}} q^3_{A}q^3}{48 \pi^3 m_{B_{s}}^5} \bar{|\mathcal{ A}|^2},  
\end{equation}  
with the variable $\xi =s/m^2_{B_{s}}$, and the $B$ meson mean lifetime $\tau_B$. Among them, $q$ is already explained in Eq.(3) and $q_{A}$ is defined as \cite{SH23}: 
\begin{equation}  
q_{A} = \frac{1}{2} \sqrt{ \left[ \left( m^2_{B_s} - m^2_{P} \right)^2 - 2 \left( m^2_{B_s} + m^2_{P} \right) s + s^2 \right]/s}.  
\end{equation}  
where $m_{P}$ is the mass of the Pseudoscalar meson, this is obtained from the momentum analysis of the final-state particle.

\section{\label{sum}Numerical results}
\subsection{\label{subsec:form}The curve results of localised CP asymmetry}
\begin{figure}[h]  
	\centering  
	\begin{minipage}{0.45\textwidth}  
		\centering  
		\includegraphics[width=\linewidth]{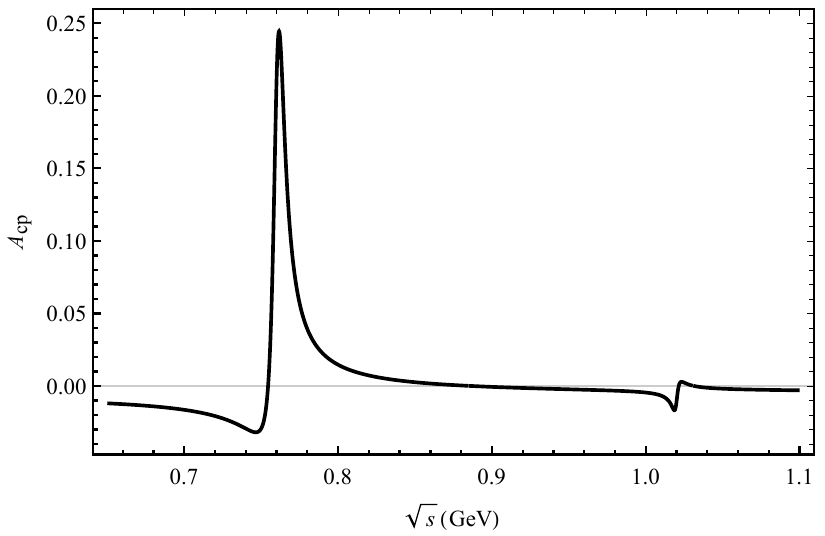}  
		\caption{The decay channel of $\bar B^{0}_{s}\rightarrow \pi^{+}\pi^{-}\pi^{0}$.}  
		\label{fig2}  
	\end{minipage}  
	\hfill  
	\begin{minipage}{0.45\textwidth}  
		\centering  
		\includegraphics[width=\linewidth]{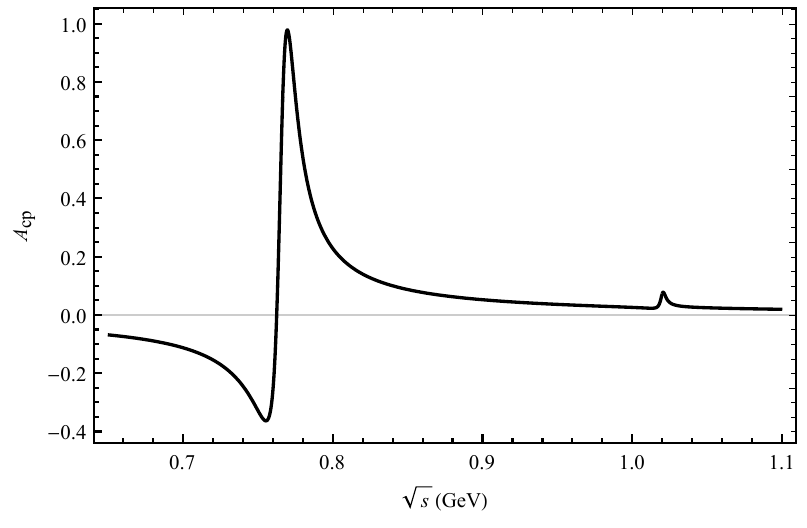}  
		\caption{The decay channel of $\bar B^{0}_{s}\rightarrow \pi^{+}\pi^{-}\bar K^{0}$.}  
		\label{fig3}  
	\end{minipage}  
\end{figure}

\begin{figure}[h]  
	\centering  
	\begin{minipage}{0.45\textwidth}  
		\centering  
		\includegraphics[width=\linewidth]{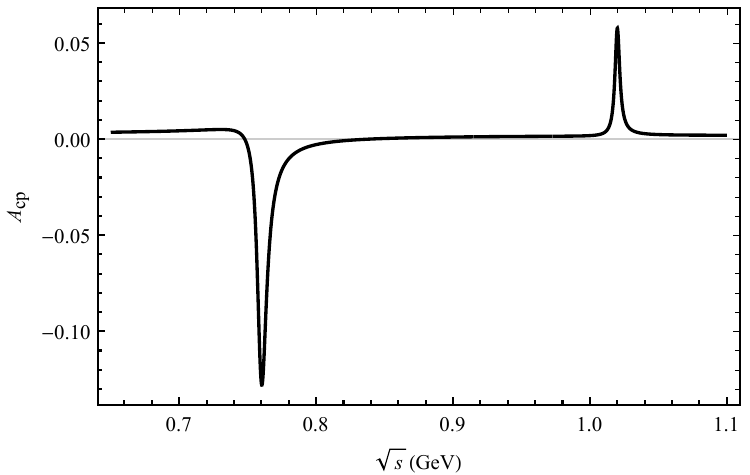}  
		\caption{The curve corresponds the decay channel of $\bar B^{0}_{s}\rightarrow \pi^{+}\pi^{-}\eta$.}  
		\label{fig4}  
	\end{minipage}  
	\hfill  
	\begin{minipage}{0.45\textwidth}  
		\centering  
		\includegraphics[width=\linewidth]{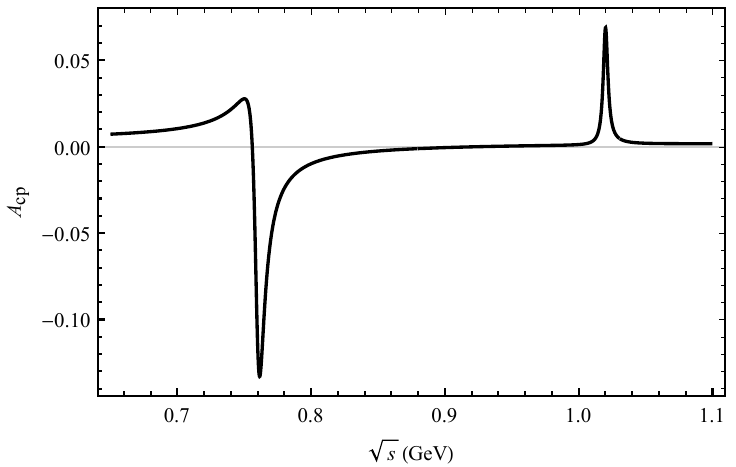}  
		\caption{The curve corresponds the decay channel of $\bar B^{0}_{s}\rightarrow \pi^{+}\pi^{-}\eta^{'}$.}  
		\label{fig5}  
	\end{minipage}  
\end{figure}

In the framework of QCDF, we calculate the decay amplitudes due to quasi-two-body decay process such as Eqs  (18)-(20). One can find that the decay amplitudes is dependent of CKM matrix elements, decay constants, form factors and Wilson coefficients for different final state particle, respectively. The strong phase $\delta$ and the absolute value $r$ of the ratio of the penguin and tree amplitudes can be calculated from the approach of QCDF which is different for different final state particles. We consider the previous results and choose the same range of threshold for the resonance effect of vector meson mixing \cite{15Li2022}. 
The Particle Data Group (PDG) data shows that the masses of $\rho$, $\omega$ and $\phi$ are approximately estimated to be $0.775$ GeV, $0.782$ GeV and $1.019$ Gev, respectively \cite{PDG}. 
We choose the region between $0.65-1.10$ GeV within our theoretical framework, where the resonance effects of mixing $\omega-\rho$ and $\phi-\rho$ can be visually observed. This is the main resonance region and decay process by $V(\rho,\omega,\phi)\rightarrow \pi^{+}\pi^{-}$ to build the plot of $A_{CP}$ as a function of $\sqrt{s}$. The change in CP asymmetry for each decay process under the influence of the resonance effect is plotted by the curve in FIG  \ref{fig2} to  \ref{fig5}.
Moreover, the invariant mass of $\pi^{+}\pi^{-}$ is shown around the mass of $\rho (\omega,\phi)$ meson, so the overall CP asymmetry is observed for numbers ranging from 0.65 to 1.10 GeV \cite{15Li2022}. The results are shown in FIG. \ref{fig2} and FIG. \ref{fig3}, which illustrate the interrelation between CP asymmetry and $\sqrt{s}$.
We see a significant peak in the CP asymmetry for the four decay modes of $\bar{B}_{s}^{0}\rightarrow \rho (\omega,\phi)\pi^{0}(\bar K^{0}, \eta,\eta^{'})\rightarrow \pi^{+}\pi^{-} \pi^{0}(\bar K^{0}, \eta,\eta^{'})$ due to the $\rho$, $\omega-\rho$ and $\phi-\rho$ resonances in FIG. \ref{fig2}, FIG. \ref{fig3}, FIG. \ref{fig4} and FIG. \ref{fig5} where $\rho$ dominates.

Since $\bar{B}^{0}_{s}\rightarrow \rho\pi^{0}$ and $\bar{B}^{0}_{s}\rightarrow \omega\pi^{0}$ have no tree level contribution for the $\bar{B}^{0}_{s}\rightarrow \pi^{+}\pi^{-}\pi^{0}$ decay mode, the generation of CP asymmetry is mainly due to weak annihilation contribution. Its CP asymmetries ranging from $24.51\%$ to $-3.30\%$ are observed in the resonance regions of $\omega-\rho$, and significant CP asymmetries ranging from $0.39\%$ to $-1.72\%$ are observed in the resonance regions of $\phi-\rho$, as shown in FIG. \ref{fig2}. 
For the $\bar{B}^{0}_{s}\rightarrow \pi^{+}\pi^{-} \bar K^{0}$ decay process, large CP asymmetries ranging from $97.41\%$ to $-36.51\%$ in the resonance regions of $\omega-\rho$, and in the resonance regions of $\phi-\rho$ the CP asymmetries ranging from $7.81\%$ to $2.15\%$, as depicted in FIG. \ref{fig3}. 
Thus, there is a large change in the CP asymmetry between the resonance regions of $\omega-\rho$ for  $\bar{B}^{0}_{s}\rightarrow \pi^{+}\pi^{-}\bar K^ {0}$ and $\bar{B}^{0}_{s}\rightarrow \pi^{+}\pi^{-}\pi^{0}$, the CP asymmetry vary slightly around the $\phi-\rho$ resonance range under QCDF.

In addition, we calculate the decay process $\bar{B}^{0}_{s}\rightarrow V \eta(\eta^{'})\rightarrow \pi^{+}\pi^{-} \eta(\eta^{'})$. 
In the SU($3$) quark representation of hadrons, the corresponding parameters are more difficult to determine due to the octet-singlet mixing, Feldmann-KrollStech (FKS) mixing scheme is adopted for $\eta-\eta^{'}$ mixing in Ref.\cite{TF98} in this paper. The parameters in the calculation can be expressed by $f_q$, $f_s$ and $\phi$, the specific amplitude forms are written in Appendix A.
The physical states of the $ \eta$ and $ \eta^{'}$ mesons consist of a mixture of flavor eigenstates, namely, $\eta_n$ and $\eta_s$. We can observe CP asymmetries ranging from $0.47\%$ to $-12.79\%$ within the resonance range of $\omega-\rho$ in the decay channel of $\bar{B}^{0}_{s}\rightarrow \pi^{+}\pi^{-} \eta$, and CP asymmetries ranging from $5.77\%$ to $0.25\%$ within the resonance range of $\phi-\rho$ in FIG. \ref{fig4}. In the case of $\bar{B}^{0}_{s}\rightarrow \pi^{+}\pi^{-} \eta^{'}$, we find significant CP asymmetries ranging from $2.76\%$ to $-13.29\%$ within the $\omega-\rho$ resonance region, and CP asymmetries ranging from $6.85\%$ to $0.14\%$ within the $\phi-\rho$ resonance region, which is depicted in FIG. \ref{fig5}. 
The CP asymmetry in the decay process $\bar{B}^{0}_{s}\rightarrow \pi^{+}\pi^{-} \eta^{'}$ exhibit a significant variation similar to that observed in the decay process of $\bar{B}^{0}_{s}\rightarrow \pi^{+}\pi^{-} \eta$, specifically within the mass resonances of $\omega-\rho$.
Similarly, the CP asymmetry of these two decay processes are also changed in the $\phi-\rho$ resonance region.

These results are obtained using the central parameter values of the CKM matrix elements. The CP asymmetry results observed during these decays hopefully provide a valuable aid in the analysis of fundamental physical phenomena for the interference of vector mesons.

\subsection{\label{subsec:form}The local integral CP asymmetry}
Due to there are no experimentally measured results under the $\rho-\omega-\phi$ mixing, we calculate the integral result of the CP asymmetry for the $\bar{B}_{s}^{0}\rightarrow \rho (\omega,\phi)\pi^{0}(\bar K^{0})\rightarrow \pi^{+}\pi^{-}\pi^{0}(\bar K^{0})$ to compare with the previous result of PQCD in this paper. We also calculate the $\bar{B}_{s}^{0}\rightarrow \rho (\omega,\phi)\eta(\eta^{'})\rightarrow \pi^{+}\pi^{-}\eta(\eta^{'})$ decay processes by integrating over the invariant masses of $m_{\pi^{+}\pi^{-}}$ in the range of $0.65$ GeV$-$$1.1$ GeV from the $\rho$, $\omega$ and $\phi$ resonance regions \cite{HYC42}. We also calculate the region of $0.75$ GeV$-$$0.82$ GeV, which is obviously CP asymmetry in the FIG. \ref{fig2} to \ref{fig5}. The results are given in Table I.

\begin{table}[!ht]
		\renewcommand
	\arraystretch {3}
	\centering %
	\renewcommand{\arraystretch}{3.0} %
	\caption{The comparison of $A^\Omega _{cp}$ from $\rho-\omega-\phi$ mixing with $V\rightarrow \pi^{+}\pi^{-}$.}
	\setlength{\tabcolsep}{1.5mm}
	\begin{tabular}{c c c c}
		\hline
		Decay channel
		& {\makecell[c]{$\rho-\omega-\phi$ mixing(PQCD)\\($0.65-1.1$ \textrm{GeV})\cite{15Li2022,PDG}}}
		& {\makecell[c]{$\rho-\omega-\phi$ mixing(QCDF)\\($0.65-1.1$ \textrm{GeV})}}
		&{\makecell[c]{$\rho-\omega-\phi$ mixing(QCDF)\\($0.75-0.82$ \textrm{GeV})}}
		\\ \hline  $\bar{B}_{s}^{0}\rightarrow \rho(\omega,\phi)\pi^{0}\rightarrow \pi^{+}\pi^{-}\pi^{0}$ &-0.008$\pm$0.002    & -0.001$\pm$0.003$\pm$0.007 & 0.014$\pm$0.006$\pm$0.011
		\\ \hline  $\bar{B}_{s}^{0}\rightarrow \rho(\omega,\phi)\bar K^{0}\rightarrow \pi^{+}\pi^{-}\bar K^{0}$   &-0.017$\pm$0.003  & 0.053$\pm$0.014$\pm$0.006   & 0.31$\pm$0.015$\pm$0.020 
		\\ \hline  $\bar{B}_{s}^{0}\rightarrow \rho(\omega,\phi)\eta\rightarrow \pi^{+}\pi^{-}\eta$    &——     & 0.001$\pm$0.006$\pm$0.001  &-0.003$\pm$0.003$\pm$0.008 
		\\ \hline  $\bar{B}_{s}^{0}\rightarrow \rho(\omega,\phi)\eta^{'} \rightarrow \pi^{+}\pi^{-}\eta^{'}$ &——   & 0.001$\pm$0.002$\pm$0.004  &-0.009$\pm$0.008$\pm$0.006 
	   \\ \hline
	\end{tabular}
\end{table}

Subsequently, we compare the previous PQCD results to see the differences between the same decay processes. It is worth noting that the decay of $\bar{B}_{s}^{0}\rightarrow \pi^{+}\pi^{-}\bar K^0$ with resonance effect has a significant effect on CP asymmetry in the same energy range under QCDF.
The CP asymmetry result of decay $\bar{B}_{s}^{0}\rightarrow \pi^{+}\pi^{-}\pi^{0}$ is significantly smaller than that of the previous PQCD approach, possibly because the amplitude contribution of the quasi-two-body decay of $\bar{B}_{s}^{0}\rightarrow \pi^{+}\pi^{-}\pi^{0}$ is mainly the penguin  level contribution and the annihilation level contribution. 
Besides, we add the two decay processes $\bar{B}_{s}^{0}\rightarrow \rho (\omega,\phi)\eta\rightarrow \pi^{+}\pi^{-}\eta$ and $\bar{B}_{s}^{0}\rightarrow \rho (\omega,\phi)\eta^{'}\rightarrow \pi^{+}\pi^{-}\eta^{'}$ and calculate that the CP asymmetry result affected by the resonance effect.
The local CP asymmetry results for all decay processes are clearly enhanced within the $0.75$ GeV$-$$0.82$ GeV region under the invariant mass region of $\omega-\rho$ mesons, especially for the decay process of  $\bar{B}_{s}^{0}\rightarrow \pi^{+}\pi^{-}\pi^{0}$ and $\bar{B}_{s}^{0}\rightarrow \pi^{+}\pi^{-}\eta^{'}$.
The CP asymmetry result of $\bar{B}_{s}^{0}\rightarrow \pi^{+}\pi^{-}\eta$ is smaller than the result produced by other decay processes during the region of $0.75$ GeV$-$$0.82$ GeV which is the mass resonances of $\omega-\rho$.

There are two uncertainties in our numerical results, the first one is caused by the uncertainty of the mixed parameter and other input parameters, and the second one is caused by parameterization of logarithmic diverging integral in the QCDF.

\subsection{\label{subsec:form}The decay branching ratio}
In view of the specific experimental data results of the decay branch ratio of the quasitwo-body decay $\bar{B}_{s}^{0}\rightarrow \rho(\omega,\phi) P\rightarrow \pi^{+}\pi^{-}P$ process without the resonance effect, we only refer to the theoretical results of the latest relevant studies.
The branching ratio of the three-body decay of $B_s$ meson has been studied more and more extensively by the PQCD approach, but the calculation results of the branching ratio under the $\rho-\omega-\phi$ mixing mechanism have not been considered, so we only use the results of the direct decay of $\bar{B}_{s}^{0}\rightarrow \rho P\rightarrow \pi^{+}\pi^{-}P$ for comparison.
As the FAT approach considered the power corrections from “chiral enhanced” term, penguin annihilation contribution and EW-penguin diagram for the $\bar{B}_{s}^{0}\rightarrow VP$ decay process, and some of their results closely matched the experimental result, so we use them for comparison.

When we calculate the branching ratio under the $\rho-\omega-\phi$ resonance effect, we have given the branching ratio without the influence of the resonance effect for the $\bar{B}_{s}^{0}\rightarrow \rho(\omega,\phi) P\rightarrow \pi^{+}\pi^{-}P$ decay process. The calculation results of the branch ratio of quasi-two-body decay based on $\rho-\omega-\phi \rightarrow \pi^{+}\pi^{-}$ are listed in Table II for easy comparison and observation.

\begin{table}[!ht]
	\renewcommand
	\arraystretch {3}
	\centering %
	\renewcommand{\arraystretch}{3.0} %
	\caption{The branching ratio $(\times10^{-6})$ of $\bar{B}_{s}^{0}\rightarrow VP \rightarrow \pi^{+}\pi^{-}P$.}
	\setlength{\tabcolsep}{6mm}
	\begin{tabular}{c c c c}
		\hline
		Decay channel
		& {\makecell[c]{PQCD approach\cite{PDG,YL17}\\$\rho\rightarrow \pi^{+}\pi^{-}$}}
		& {\makecell[c]{FAT approach\cite{PDG,SH23}\\$\rho\rightarrow \pi^{+}\pi^{-}$}}
		& {\makecell[c]{QCDF approach\\\makecell[c]{$\rho\rightarrow \pi^{+}\pi^{-}$\\$\rho-\omega-\phi \rightarrow \pi^{+}\pi^{-}$}}}
		\\ \hline  $\bar{B}_{s}^{0}\rightarrow \pi^{+}\pi^{-}\pi^{0}$ &0.35 $^{+0.06}_{-0.05}$ $\pm$0.01$\pm$0.00  & 0.35$\pm$0.05$\pm$0.01$\pm$0.03  &{\makecell[c]{0.16$\pm$0.03$\pm$0.02$\pm$0.04\\0.15$\pm$0.04$\pm$0.09$\pm$0.01}}
		\\ \hline  $\bar{B}_{s}^{0}\rightarrow \pi^{+}\pi^{-}\bar K^{0}$ &  0.21 $^{+0.05}_{-0.01}$ $^{+0.01}_{-0.00}$ $^{+0.01}_{-0.00}$ &1.55$\pm$1.10$\pm$0.31$\pm$0.02 & {\makecell[c]{0.12$\pm$0.05$\pm$0.09$\pm$0.04\\0.08$\pm$0.06$\pm$0.02$\pm$0.00}}
		\\ \hline  $\bar{B}_{s}^{0}\rightarrow \pi^{+}\pi^{-}\eta$ &   0.10 $^{+0.04}_{-0.02}$ $\pm$0.00$\pm$0.00
		&0.11$\pm$0.02$\pm$0.02$\pm$0.03   &{\makecell[c]{0.32$\pm$0.04$\pm$0.03$\pm$0.03\\0.30$\pm$0.01$\pm$0.04$\pm$0.05}}
		\\ \hline  $\bar{B}_{s}^{0} \rightarrow \pi^{+}\pi^{-}\eta^{'}$ & 0.23 $^{+0.08}_{-0.06}$ $^{+0.00}_{-0.01}$ $\pm$0.00
		&0.34$\pm$0.07$\pm$0.05$\pm$0.01   &{\makecell[c]{0.22$\pm$0.04$\pm$0.05$\pm$0.02\\0.20$\pm$0.06$\pm$0.03$\pm$0.01}}
		\\ \hline
	\end{tabular}
\end{table}
By comparison, it can be seen that in the direct decay process of $\bar{B}_{s}^{0}\rightarrow \rho  P\rightarrow\pi^{+}\pi^{-}P$, the decay branching ratio calculated by the three approachs is of one order of magnitude. For the $\bar{B}_{s}^{0}\rightarrow \pi^{+}\pi^{-}\pi^{0}$ process, the calculated result of QCDF approach is lower than that of the other two processes because the decay amplitude of the process is dominated by the annihilation graph. For the decay process of $\bar{B}_{s}^{0}\rightarrow \pi^{+}\pi^{-}\bar K^{0}$, the results of the QCDF approach are close to those of the PQCD approach, but the details of the FAT approach are larger, which may be caused by the shortage of nonperturbative contribution and $1/m_b$ power corrections of the FAT approach. As a result of the direct decay process of $\bar{B}_{s}^{0}\rightarrow \pi^{+}\pi^{-}\eta$,
The QCDF result is larger than the other two processes, which may be due to the error introduced by the form factor itself of the QCDF approach.
In the decay process of $\bar{B}_{s}^{0} \rightarrow \pi^{+}\pi^{-}\eta^{'}$, the results of QCDF and PQCD are consistent, and the difference between the results of QCDF and the results of FAT approach is also very small. Due to the difference in the treatment of intermediate virtual particles in FAT approach, the results may be biased.

Above all, we calculate the decay branching ratio of these four decay processes under the $\rho-\omega-\phi$ resonance effect. We find that the results of each decay branching ratio are suppressed, especially for the $\bar{B}_{s}^{0}\rightarrow \pi^{+}\pi^{-}\bar K^{0}$ decay process. We think that the mixing of the intermediate state meson $\rho-\omega-\phi$ will depress the decay branching ratio.
Since intermediate virtual particles such as $\rho-\omega$ can not be effectively distinguished experimentally, it is necessary to consider the effects of mixing of intermediate virtual particles in future research.

Similarly, errors are also taken into account in the calculation. The first error is caused by CKM matrix elements, form factors and decay constants, the second error is caused by QCDF approach itself in the calculation process, and the third error is caused by mixed parameters.

\section{\label{sum}SUMMARY}
Our results show that CP asymmetries and decay branching ratios have obvious changes due to the resonance effect of $V\rightarrow \pi^{+}\pi^{-}$ $(V=\rho,\omega,\phi)$ in the $B \rightarrow \pi^{+}\pi^{-} P$ decay modes when the invariant mass of $\pi^{+}\pi^{-} $ is close to the $\omega-\rho$ and $\phi-\rho$ resonance regions within the framework of QCDF. 

The three-body decay process is efficiently calculated by using quasi-two-body chain decay. Taking $B\rightarrow RP_3$ as an example, the intermediate resonance state $R$ decays into two hadrons $P_1$ and $P_2$, while $P_3$ is the other hadron. This process can be decomposed using the narrow width approximation as $\mathcal{B} (B\rightarrow RP_3\rightarrow P_1 P_2 P_3)=\mathcal{B} (B\rightarrow RP_3)\mathcal{B} (B\rightarrow P_1 P_2)$. 
In small widths, the effects of $\omega$ and $\phi$ can be ignored in quasi-two-body chain decay. As a measure of the degree of approximation of $\Gamma (B\rightarrow RP_3) \mathcal{B} (B\rightarrow P_1 P_2)=\eta_{R}\Gamma(B\rightarrow RP_3\rightarrow P_1 P_2 P_3)$ , the parameter $\eta_{R}$ is introduced \cite{46HY}.  
The integral of the invariant mass $m_{\pi^+ \pi^-}$ is considered in the calculation. Due to the attenuation amplitude has a Breit-Wigner shape and depends on the parameter of the invariant mass $m_{\pi^+ \pi^-}$. 
In the present manuscript, the effect of this correction is ignored in view of the range of accuracy. This level of correction is about 7$\%$ and is also a source of error in our results \cite{HYC42}.

In the local integration results (in Table I), 
the local CP asymmetry associated with $B \rightarrow \rho(\omega,\phi)P \rightarrow \pi^{+}\pi^{-}P$ can be found by calculating the specific phase space region. Due to the interference of $\rho-\omega-\phi$ caused by the breaking of isospin, the resonance contribution of $\omega-\rho$ and $\phi-\rho$ can produce a new strong phase, which has a great influence on the CP asymmetry of the $B \rightarrow \pi^{+}\pi^{-}P$ decay mode.
It is evident that for the $\bar{B}_{s}^{0} \rightarrow \pi^{+}\pi^{-} \pi^{0}$ process, the result obtained by the QCDF approach is significantly smaller than that obtained by the PQCD approach. This is because the amplitude contribution of the $\bar{B}_{s}^{0} \rightarrow \pi^{+}\pi^{-} \pi^{0}$ process is dominated by the penguin and annihilation contribution. On the contrary, the CP asymmetry result of the $\bar{B}_{s}^{0} \rightarrow \pi^{+}\pi^{-}\bar K^{0}$ process is much larger due to the obvious tree and penguin level contributions in this process, while the uncertainty of form factor and decay constant also add some errors to the results.
For the newly added decay processes of $\bar{B}_{s}^{0}\rightarrow \pi^{+}\pi^{-}\eta$ and $\bar{B}_{s}^{0}\rightarrow \pi^{+}\pi^{-}\eta^{'}$, when the threshold interval is 0.75-0.82 GeV, the local integral result becomes significantly larger, indicating that the CP asymmetry of $\bar{B}_{s}^{0}\rightarrow \pi^{+}\pi^{-}\eta$ and $\bar{B}_{s}^{0}\rightarrow \pi^{+}\pi^{-}\eta^{'}$ decays mainly occurs in the mixing region of $\omega-\rho$ since $\rho$ is dominant.

In the process of calculating the decay branching ratio (in Table II), it can also be found that the decay branching ratio will be smaller when considering the resonance effect. The results with and without resonance effects are compared with PQCD approach and FAT approach. In general, due to the lack of non-perturbation contribution and $1/m_b$ power correction, the attenuation amplitude and phase extracted from the experimental data by the FAT approach are larger than that by the PQCD approach or the QCDF approach, resulting in a larger decay branching ratio result. The PQCD approach only considers the decay branching ratio of $\bar{B}_{s}^{0}\rightarrow \rho P\rightarrow \pi^{+}\pi^{-}P$ in the direct decay process, which is close to our result without considering the resonance effect. We also find that the decay branching ratio is lower when considering resonance effects, especially for the $\bar{B}_{s}^{0} \rightarrow \pi^{+}\pi^{-}\bar K^{0}$ process. In addition, we also give the error term in the calculation results.

Generally, researchers can reconstruct the intermediate virtual particles $\rho$, $\omega$ and $\phi$ from the final state mesons of $\pi^{+}\pi^{-}$ to measure the decay branching ratio and the predicted CP asymmetry in experiment. However, it is difficult to distinguish and analyze the effect of intermediate mesons $\rho$ and $\omega$ on the final state particle production during the experiment. Therefore, it is necessary to consider the resonance effect. We hope that these work can provide support for the experimental research.

\section{Acknowledgments}
This work was supported by National Natural Science Foundation of China (Project No.11805153) 
and Natural Science Foundation of Henan Province (Project No.252300420319). 

%\newpage

\end{spacing}

\begin{appendix}
	\section{\label{sum}DECAY AMPLITUDE}
	\begin{eqnarray}
	\begin{array}{c}
	\mathcal{M}\left(\bar{B}^{0}_{s}\rightarrow \rho(\rho \rightarrow \pi^{+}\pi^{-}) \bar{K}^{0} \right)=$$\sum_{q =u,c}$$ {\frac{G_Fg_{\rho \pi^{+}\pi^{-}}m_{\rho}\epsilon(\lambda) \cdot p_{K}}{s_{\rho}}}\left\{V_{ub}V_{ud}^{*} f_{\rho}a_2 +V_{tb}V_{td}^{*}\left[f_{K}A_{0}^{B\rightarrow \rho}\right.\right.\\
	\\
	\left.\left.\left( a_4-\frac{1}{2}a_{10}-\frac{3}{2}a_{7}-\frac{3}{2}a_{9}\right)
	+\frac{1}{2m_{\rho}\epsilon(\lambda) \cdot p_{K}}f_{B_s} f_{K} f_{\rho}(b_{3}(\rho,K)-\frac{1}{2}b_{3}^{EW}(\rho,K))\right]\right\},
	\end{array}
	\end{eqnarray}
	\begin{eqnarray}
	\begin{array}{c}
	\mathcal{M}\left(\bar{B}^{0}_{s}\rightarrow \omega(\omega \rightarrow \pi^{+}\pi^{-}) \bar{K}^{0} \right)=$$\sum_{q =u,c}$$ {\frac{G_Fg_{\omega\pi^{+}\pi^{-}\pi^0}m_{\omega}\epsilon(\lambda) \cdot p_{K}}{s_{\omega}}}\left\{V_{ub}V_{ud}^{*} f_{\omega}F_{1}^{B^{0}_{s}\rightarrow K}a_2\right.\\
	\\
	\left.-V_{tb}V_{td}^{*}\left[f_{\omega}F_{1}^{B^{0}_{s}\rightarrow K}\left( 2a_3+2a_5+a_4-\frac{1}{2}a_{10}+\frac{1}{2}a_{7}+\frac{1}{2}a_{9} \right)\right.\right.\\
	\\
	\left.\left.-\frac{1}{2m_{\omega}\epsilon(\lambda) \cdot p_{K}}f_{B_s} f_{K} f_{\omega}(b_{3}(\omega,K)-\frac{1}{2}b_{3}^{EW}(\omega,K))\right]\right\},
	\end{array}
	\end{eqnarray}
	\begin{eqnarray}
	\begin{array}{c}
	\mathcal{M}\left(\bar{B}^{0}_{s}\rightarrow \phi(\phi \rightarrow \pi^{+}\pi^{-})\bar K^0 \right) =$$\sum_{q =u,c}$$ {\frac{G_F g_{\phi  \pi^{+}\pi^{-} }m_{\phi}\epsilon(\lambda) \cdot p_{K}}{s_\phi}}\left\{V_{tb}V_{ts}^{*}\left[-\sqrt{2}f_{\phi}F_{1}^{B^{0}_{s} \rightarrow K}\right.\right.\\
	\\
	\left.\left.(a_3+a_5-\frac{1}{2}a_7-\frac{1}{2}a_9)-\sqrt{2}f_{K}A_{0}^{B^{0}_{s} \rightarrow \phi}(a_4-\frac{1}{2}a_{10}-a_6O_1+\frac{1}{2}a_8O_1)\right.\right.\\
	\\
	\left.\left.-\frac{1}{\sqrt{2}m_{\phi}\epsilon(\lambda) \cdot p_{K}}f_{B_s} f_{K} f_{\phi}(b_{3}(K,\phi)-\frac{1}{2}b_{3}^{EW}(K,\phi))\right]\right\},
	\end{array}
	\end{eqnarray}
	\begin{eqnarray}
	\begin{array}{c}
	\mathcal{M}\left(\bar{B}_{s}^{0}\rightarrow \rho(\rho \rightarrow \pi^{+}\pi^{-})  \eta^{(')} \right)=$$\sum_{q =u,c}$$ {\frac{G_F g_{\rho \pi^{+}\pi^{-}}m_{\rho}\epsilon(\lambda) \cdot p_{\eta^{(')}}}{s_\rho}} \\
	\\
	\left\{V_{ub}V_{us}^{*}\left[f_{\rho}F_{1}^{B_{s}^{0}\rightarrow \eta^{(')}}a_2+\frac{1}{2m_{\rho}\epsilon(\lambda) \cdot p_{\eta^{(')}}}f_{B_s}f_{\rho} f^u_{\eta^{(')}} (b_{1}(\eta^{(')},\rho)+b_{1}(\rho,\eta^{(')}))\right]\right.\\
	\\
	\left. -V_{tb}V_{ts}^{*}\left[f_{\rho}F_{1}^{B^{0}_{s}\rightarrow \eta^{(')}} \left(\frac{3}{2}a_7+\frac{3}{2}a_9\right)+\frac{1}{2m_{\rho}\epsilon(\lambda) \cdot p_{\eta^{(')}}}f_{B_s}f_{\rho} f^u_{\eta^{(')}}(\frac{3}{2}b_{4}^{EW}(\eta^{(')},\rho)+\frac{3}{2}b_{4}^{EW}(\rho,\eta^{(')})) \right]\right\},
	\end{array}
	\end{eqnarray}
	\begin{eqnarray}
	\begin{array}{c}
	\mathcal{M}\left(\bar{B}_{s}^{0}\rightarrow \omega(\omega \rightarrow \pi^{+}\pi^{-})  \eta^{(')} \right) =$$\sum_{q =u,c}$$ {\frac{G_F g_{\omega  \pi^{+}\pi^{-}
				m_{\omega}\epsilon(\lambda) \cdot p_{\eta^{(')}}}}{s_\omega}} \left\{V_{ub}V_{us}^{*}\left[f_{\omega}F_{1}^{B_{s}^{0}\rightarrow \eta^{(')}}a_2\right.\right. \\
	\\
	\left.\left.+\frac{1}{2m_{\omega}\epsilon(\lambda) \cdot p_{\eta^{(')}}}f_{B_s}f_{\omega} f^u_{\eta^{(')}} (b_{1}(\eta^{(')},\omega)+b_{1}(\omega,\eta^{(')}))\right]-V_{tb}V_{ts}^{*}\left[f_{\omega}F_{1}^{B^{0}_{s}\rightarrow \eta^{(')}}\left(2a_3+2a_5+\frac{1}{2}a_7 \right.\right.\right.\\
	\\
	\left.\left. \left.+\frac{1}{2}a_9\right)+\frac{1}{2m_{\omega}\epsilon(\lambda) \cdot p_{\eta^{(')}}}f_{B_s}f_{\omega}f^u_{\eta^{(')}}(b_{4}(\eta^{(')},\omega)+b_{4}(\omega,\eta^{(')})+\frac{1}{2}b_{4}^{EW}(\eta^{(')},\omega)+\frac{1}{2}b_{4}^{EW}(\omega,\eta^{(')})) \right]\right\},
	\end{array}
	\end{eqnarray}
	\begin{eqnarray}
	\begin{array}{c}
	\mathcal{M}\left(\bar{B}_{s}^{0}\rightarrow \phi(\phi \rightarrow \pi^{+}\pi^{-})  \eta^{(')} \right)=$$\sum_{q =u,c}$$ {\frac{G_Fg_{\phi  \pi^{+}\pi^{-}m_{\phi}\epsilon(\lambda) \cdot p_{\eta^{(')}} }}{s_{\phi}}}\left\{V_{ub}V_{us}^{*}f^u_{\eta^{(')}} A_0^{B^0_s\rightarrow \phi}a_2-V_{tb}V_{ts}^{*} \right.\\
	\\
	\left.\left[ f^u_{\eta^{(')}} A_0^{B^0_s\rightarrow \phi}\left\{2a_{3}-2a_{5}-\frac{1}{2}a_7+\frac{1}{2}a_9+\left\{a_3-a_5+a_4-\frac{1}{2}a_{10}+\frac{1}{2}a_7-\frac{1}{2}a_9-(a_6-\frac{1}{2}a_8)O_2^{(')}(1-\frac{f^u_{\eta^{(')}}}{f^s_{\eta^{(')}}})\right\}\right.\right.\right.\\
	\\
	\left.\left.\left.\frac{f^s_{\eta^{(')}}}{f^u_{\eta^{(')}}}\right\}
	+f_{\phi}F_1^{B_0^s\rightarrow \eta^{(')}}(a_3+a_5+a_4-\frac{1}{2}a_{10}-\frac{1}{2}a_9-\frac{1}{2}a_7)+\frac{1}{\sqrt 2m_{\phi}\epsilon(\lambda) \cdot p_{\eta^{(')}} }f_{B_s}f_{\phi}f^s_{\eta^{(')}} (b_{3}(\eta^{(')},\phi)\right.\right.\\
	\\
	\left.\left.
	+b_{3}(\phi,\eta^{(')})-\frac{1}{2}b_{3}^{EW}(\eta^{(')},\phi)-\frac{1}{2}b_{3}^{EW}(\phi,\eta^{(')})+b_{4}(\eta^{(')},\phi)+b_{4}(\phi,\eta^{(')})
	-\frac{1}{2}b_{4}^{EW}(\eta^{(')},\phi)-\frac{1}{2}b_{4}^{EW}(\phi,\eta^{(')}))\right]\right\},
	\end{array}
	\end{eqnarray}
The form of coupling constant $g$ and parameter $O$ in the paper is as follows:
\begin{eqnarray}
\begin{array}{c}
g_{\rho^0\pi\pi}^{2}=\frac{48\pi}{(1-\frac{4m_{\pi}^2}{m_{\rho}^2})^{3/2}}\times \frac{\Gamma_{\rho^0\rightarrow \pi^+\pi^-}}{m_\rho},\quad
O_1=\frac{2m_{K^{0}}^{2}}{(m_b+m_s)(m_d+m_s)},\quad
O_2^{(')}=\frac{2m_{\eta^{(')}}^{2}}{(m_b+m_s)(m_s+m_s)}.
\end{array}
\end{eqnarray}	
	
	\section{\label{sum}INPUT PARAMETER}
	
	\begin{table}[!ht]
		\renewcommand
		\arraystretch {3}
		\centering %
		\renewcommand{\arraystretch}{3.0} %
		\caption{Input Parameter Value (GeV) \cite{2lu2021,PDG,15Li2022,JC21,XL05}.}
		\setlength{\tabcolsep}{1.2mm}
		\begin{tabular}{|c|c|c|c|}
			\hline
				$\lambda_{CKM}=0.22650\pm0.00048$ &$A_{CKM}=0.790^{+0.017}_{-0.012}$ & {$\bar \rho_{_{CKM}}=0.141^{+0.016}_{-0.017}$}  &{$\bar \eta_{_{CKM}}=0.357\pm0.01$ }    
			\\ 	\hline
			$m_{B^0_s}=5.36692\pm0.00010$ &$m_{\rho}=0.77526\pm0.00023$ & {$m_{\omega}=0.78266\pm0.00013$}  &{$m_{\phi}=1.019461\pm0.000016$ }    
			\\ 	\hline
			$f_{\omega}=0.192\pm0.010$	& {$f_{\rho}=0.213\pm0.011$} & $f_{\phi}=0.225\pm0.011$ &$f_{B_s}=0.23\pm0.03$  
			\\  \hline  {$f_{\pi}=0.130\pm0.001$} & $f_{K}=0.155\pm0.004$  & $f^q_{\eta^{(')}}=(1.07\pm0.02)f_\pi $ & $f^s_{\eta^{(')}}=(1.34\pm0.06)f_\pi $ 
			\\ \hline   $A_{0}^{B^0_s\rightarrow \phi}=0.272$ &$F_{1}^{B^0_s\rightarrow K}=0.31$ & $F_{1}^{B^0_s\rightarrow \eta_{s\bar s}}=0.335$ &  $F_{1}^{B^0_s\rightarrow \eta^{(')}_{s\bar s}}=0.282$
			\\ \hline  
		\end{tabular}
	\end{table}

\end{appendix}

\begin{thebibliography}{}
\bibitem{N63} N. Cabibbo, Physical Review Letters {\bf10}, 531 (1963).
\bibitem{M73} M. Kobayashi and T. Maskawa, Progress of Theoretical Physics {\bf49}, 652 (1973).
\bibitem{T92} T.M. Yan, H.Y. Cheng, C.Y. Cheung, G.L. Lin, Y.C. Lin, and H.L. Yu, Physical Review D {\bf46}, 1148 (1992).

\bibitem{JJQ} J.J. Qi, X.H. Guo, Z.Y. Wang, Z.H. Zhang, C. Wang, Physical Review D {\bf99}, 076010 (2019).
\bibitem{SZ} S.H. Zhou, R.H. Li, Z.Y. Wei, C.D. Lü, Physical Review D  {\bf104}, 116012 (2021).

\bibitem{15Li2022} Ahmed. Ali, Gustav. Kramer, Physical Review D {\bf76}, 074018 (2007).


\bibitem{E01}R. Aaij et al. (LHCb Collaboration), Physical Review Letters {\bf112}, 011801
(2014).
\bibitem{E02} R. Aaij et al. (LHCb Collaboration), Physical Review D {\bf90}, 112004 (2014).
\bibitem{E03} J. P. Lees et al. (BABAR Collaboration), Physical Review D {\bf96},072001 (2017).


\bibitem{M85} M. Wirbel, B. Stech, M. Bauer, Zeitschrift für Physik C {\bf29}, 637 (1985).
\bibitem{M87} M. Bauer, B. Stech, M. Wirbel, Zeitschrift für Physik C {\bf34}, 103 (1987).

\bibitem{N1999} M. Beneke, G. Buchalla, M. Neubert, C.T. Sachrajda, Physical Review Letters, {\bf83}, 1914 (1999).
\bibitem{N2007} M. Beneke, J. Rohrer, D. Yang, Nuclear Physics B {\bf774}, 64 (2007).
\bibitem{H2008} H.Y. Cheng, K.C. Yang, Physical Review D {\bf78}, 094001 (2008).
\bibitem{Y01} Y.Y. Keum, H.N. Li, A.I. Sanda, Physical Review D {\bf63}, 054008 (2001).
\bibitem{YY01} Y.Y. Keum, H.N. Li, Physical Review D {\bf63}, 074006 (2001).

\bibitem{CD01} C.D. Lu, K. Ukai, M.Z. Yang, Physical Review D {\bf63}, 074009 (2001).
\bibitem{C01} C.W. Bauer, D. Pirjol, I.W. Stewart, Physical Review Letters {\bf87}, 201806 (2001).
\bibitem{C02} C.W. Bauer, D. Pirjol, I.W. Stewart, Physical Review D {\bf65}, 054022 (2002).

\bibitem{HY20}H.Y. Cheng, C.W. Chiang, A.L. Kuo, Physical Review D {\bf91}, 014011 (2015).
\bibitem{XG21} X.G. He, Y.J. Shi, W. Wang, European Physical Journal C {\bf80}, 359 (2020).
\bibitem{XG22} X.G. He, W. Wang, Chinese Physics C {\bf42}, 103108 (2018).


\bibitem{NM67} N.M. Kroll, T.D. Lee, B. Zumino, Physical Review  {\bf157}, 1376 (1967).
\bibitem{PDG} S. Navas, et al., (Particle Data Group) Physical Review D {\bf110}, 030001 (2024).

\bibitem{2lu2022} G. Lü, Y.L. Zhao, L.C. Liu, X.H. Guo, Chinese Physics C {\bf46}, 113101 (2022).
\bibitem{MDA1979} M. Bander, D. Silverman, A. Soni, Physical Review Letters {\bf43}, 242 (1979).

\bibitem{P2017} D.S. Shi, G. Lü, Y.L. Zhao, N. Wang, X.H. Guo, European Physical Journal C {\bf83}, 345 (2023).
\bibitem{SH23} S.H. Zhou, X.X. Hai, R.H. Li, C.D. Lü, Physical Review D {\bf107}, 116023 (2023).

\bibitem{2lu2021} X.L. Yuan, G. Lü, N. Wang, L.Y. Zhang, X.H. Guo, Chinese Physics C {\bf47}, 113101 (2023).
\bibitem{luG2024} G. Lü, C.C. Zhang, Y.L. Zhao, L.Y. Zhang, Chinese Physics C {\bf48}, 013103 (2024).
\bibitem{JM47} J. M. Blatt and V. F. Weisskopf, Theoretical Nuclear Physics (Springer, New York, 1952).
\bibitem{RA46} R. Aaij et al. (LHCb Collaboration), Phys. Rev. D {\bf 101}, 012006 (2020).
\bibitem{J2003} P.D. Ruiz-Femen´ıa, A. Pich, J. Portol´es Journal of High Energy Physics {\bf07}, 003 (2003).

\bibitem{MN2000} M.N. Achasov,  V.M. Aulchenko,  A.V. Berdyugin, et.al., Nuclear Physics B {\bf569}, 158 (2000).

\bibitem{P2009} C.E. Wolfe, K.Maltman, Physical Review D {\bf80}, 114024 (2009).
\bibitem{P2011} C.E. Wolfe, K.Maltman, Physical Review D {\bf83}, 077301 (2011).
\bibitem{N2003} M. Beneke, M. Neubert, Nuclear Physics B {\bf675}, 333 (2003).
\bibitem{G1996} G. Buchalla, A.J. Buras, and M.E. Lautenbacher, Reviews of Modern Physics {\bf68}, 1125 (1996).
\bibitem{JF2002} D.S. Du, H.J. Gong, J.F. Sun, D.H. Yang and G.H. Zhu, Physical Review D {\bf65}, 094025 (2002).

\bibitem{JF20031} D.S. Du, J. F. Sun, D. H. Yang and G. H. Zhu, Physical Review D {\bf67}, 014023 (2003).
\bibitem{P2022} H.Q. Liang, X.Q. Yu, Physical Review D {\bf105}, 096018 (2022).
\bibitem{HYC42} H.Y. Cheng, C.W. Chiang, C.K. Chua, Physics Letters B {\bf813}, 136058 (2021).
\bibitem{PRD2000} M.Z. Yang, Y.D. Yang, Physical Review D {\bf62}, 114019 (2000).
\bibitem{J2016} J.F. Sun, G.H. Zhu and D.S. Du, Physical Review D {\bf68}, 054003 (2003).

\bibitem{PR2000} T. Muta, A. Sugamoto,  Y.D. Yang, Physical Review D {\bf62}, 094020 (2000).
\bibitem{S2009} S. Leupold, M.F.M. Lutz, the European Physical Journal A {\bf39}, 205 (2009).
\bibitem{J2020} C. Wang, R.W. Wang, X.W. Kang, et al., Physical Review D {\bf99}, 074017 (2019).
\bibitem{EP2013} Z.H. Zhang, Y.D. Yang, X.H. Guo, G. Lü and A. Wiranata, the European Physical Journal C {\bf73}, 2555 (2013).
\bibitem{R2014} Z.H. Zhang, X.H. Guo, Y.D. Yang, Physical Review D {\bf87}, 076007 (2013).

\bibitem{WZWG2015} C. Wang, Z.H. Zhang, Z.Y. Wang, X.H. Guo, European Physical Journal C {\bf75}, 1 (2015).
\bibitem{WC2017} C. Wang, L.L. Liu, X.H. Guo. Physical Review D {\bf 96}, 056002 (2017).


\bibitem{JC21} J. Chai, S. Cheng and W. F. Wang, Physical Review D {\bf103}, 096016 (2021).
\bibitem{OL} O. Leitner, X.H. Guo, A. W. Thomas, Phys. G: Nucl. Part. Phys. {\bf31}, 199 116023 (2005).
\bibitem{BA51} B. El-Bennich, A. Furman, et al., Physical Review D {\bf79}, 094005 (2009).

\bibitem{TF98} T. Feldmann, P. Kroll, B.Stech, Physical Review D {\bf 58}, 114006 (1998).
\bibitem{YL17} Y. Li, A.J. Ma, W.F. Wang,  Z.J. Xiao, Physical Review D {\bf 95}, 056008 (2017).

\bibitem{46HY} H.Y. Cheng, C.W. Chiang, C.K. Chua, Physical Review D {\bf103}, 036017 (2021).
\bibitem{XL05} X.L. Yuan, G. Lü, N. Wang, C. Wang, Chinese Physics C  {\bf 49}, 023101 (2025).



\end{thebibliography}
\end{document}